\documentclass[12pt,oneside, a4paper]{article}
\pdfoutput=1

\ifx\pdfoutput\undefined
\usepackage[dvips,bookmarks=false]{hyperref}	
\else
\usepackage{hyperref}	
\fi
\hypersetup{colorlinks,bookmarksopen,bookmarksnumbered,citecolor=blue,
linkcolor=black,pdfstartview=FitH,urlcolor=blue}

\usepackage{graphicx}
\usepackage{subfigure}
\usepackage{amssymb}
\usepackage{cite}
\usepackage{bm}
\usepackage{amsmath,amsthm}

\newcommand{\capdef}{}
\newcommand{\mycaption}[2][\capdef]{\renewcommand{\capdef}{#2}%
       \caption[#1]{{\footnotesize #2}}}

\newcommand{\be}{\begin{equation}}
\newcommand{\ee}{\end{equation}}

\newcommand{\deltaCP}{\ensuremath{\delta_{\rm CP}}}

\begin{document}

\begin{titlepage}

\begin{center}

\vspace{1cm}
{\Large\bf On the determination of the leptonic CP phase}
\vspace{1cm}

\renewcommand{\thefootnote}{\fnsymbol{footnote}}
{\bf Jessica Elevant}\footnote[1]{Email: jessica.elevant AT fysik.su.se}, 
{\bf Thomas Schwetz}\footnote[2]{Now at: Institut f\"ur Kernphysik,
Karlsruhe Institute of Technology (KIT),
76021 Karlsruhe, Germany. Email: schwetz AT kit.edu},
\vspace{5mm}

{\it%
{Oskar Klein Centre for Cosmoparticle Physics,\\ 
Department of Physics, Stockholm University, SE-10691 Stockholm, Sweden}}

\vspace{8mm} 

\abstract{The combination of data from long-baseline and reactor
  oscillation experiments leads to a preference of the leptonic CP
  phase $\deltaCP$ in the range between $\pi$ and $2\pi$. We study the
  statistical significance of this hint by performing a Monte Carlo
  simulation of the relevant data. We find that the distribution of
  the standard test statistic used to derive confidence intervals for
  $\deltaCP$ is highly non-Gaussian and depends on the unknown true
  values of $\theta_{23}$ and the neutrino mass ordering. Values of
  $\deltaCP$ around $\pi/2$ are disfavored at between $2\sigma$ and
  $3\sigma$, depending on the unknown true values of $\theta_{23}$ and
  the mass ordering. Typically the standard $\chi^2$ approximation
  leads to over-coverage of the confidence intervals for
  $\deltaCP$. For the 2-dimensional confidence region in the
  ($\deltaCP,\theta_{23}$) plane the usual $\chi^2$ approximation is
  better justified. The 2-dimensional region does not include the
  value $\deltaCP = \pi/2$ up to the 
  86.3\% (89.2\%)~CL 
  assuming a true normal (inverted) mass ordering.  Furthermore, we study the
  sensitivity to $\deltaCP$ and $\theta_{23}$ of an increased exposure
  of the T2K experiment, roughly a factor 12 larger than the current
  exposure and including also anti-neutrino data. Also in this case
  deviations from Gaussianity may be significant, especially if the
  mass ordering is unknown.  }

\end{center}
\end{titlepage}

\renewcommand{\thefootnote}{\arabic{footnote}}
\setcounter{footnote}{0}

\setcounter{page}{2}

\tableofcontents

\section{Introduction}

Thanks to a
beautiful series of neutrino oscillation experiments 
\cite{Fukuda:1998mi, Wendell:2010md, 
Ahmad:2002jz, Araki:2004mb, 
Adamson:2008zt,  
Adamson:2013ue,  
Adamson:2013whj, 
Abe:2012tg, Abe:2014bwa, 
Ahn:2012nd, 
An:2012eh, 
An:2013zwz, 
Abe:2013hdq, 
Abe:2014ugx, 
Abe:2015awa} 
we have now a clear picture of the mixing pattern in the lepton
sector.  Neutrino oscillations depend on two neutrino mass-squared
differences $\Delta m^2_{21}, \Delta m^2_{32}$ (with $\Delta m^2_{ij}
\equiv m^2_i - m^2_j$), three mixing angles, $\theta_{12},
\theta_{23}, \theta_{13}$, and one complex phase $\deltaCP$, where we
adopt the standard parameterization for the leptonic mixing matrix
\cite{Agashe:2014kda}. Out of those six parameters the three mixing
angels and the two mass-squared differences are well determined by
global data~\cite{Gonzalez-Garcia:2014bfa, Capozzi:2013csa,
  Forero:2014bxa}, up to the sign of $\Delta m^2_{32}$ which
parametrizes two possible orderings of the neutrino mass states,
normal ordering (NO) versus inverted ordering (IO).

One of the ultimate goals of neutrino oscillation physics is to
determine the complex phase $\deltaCP$. Values of $\deltaCP$ different
from zero and $\pi$ imply CP violation in the lepton sector
\cite{Cabibbo:1977nk, Bilenky:1980cx, Barger:1980jm}, see
refs.~\cite{Nunokawa:2007qh, Branco:2011zb} for reviews. Determining
$\deltaCP$ and possibly establishing CP violation with reasonable
precision is a formidable task which most likely will require
high-intensity neutrino beams beyond the current generation of
experiments. Nevertheless, already with current experiments, some
first hints on a preferred range of $\deltaCP$ may be obtained at a
modest confidence level, see for instance \cite{Huber:2009cw,
 Prakash:2012az, Agarwalla:2012bv, Machado:2013kya, Ghosh:2015ena} for
estimates. Indeed, currently available global data seem to indicate a
slight preference for the range $\pi < \deltaCP < 2\pi$ compared to $0
< \deltaCP < \pi$~\cite{Gonzalez-Garcia:2014bfa, Capozzi:2013csa,
  Forero:2014bxa}. This hint emerges mostly from the combination of
the $\nu_\mu \to \nu_e$ observation from the long-baseline experiments
T2K~\cite{Abe:2013hdq} and MINOS~\cite{Adamson:2013ue} with the
determination of the mixing angle $\theta_{13}$ by reactor experiments
DayaBay~\cite{An:2013zwz}, RENO \cite{Ahn:2012nd}, and
DoubleChooz~\cite{Abe:2014bwa}.

Assessing the significance of this hint is a non-trivial
task. Standard statistical tools usually employed in global fits
of neutrino oscillation data are likely to fail for the determination
of $\deltaCP$. The reason is that several conditions for Wilks
theorem~\cite{wilks} are violated in this case: statistics is low, the
sensitivity of the data to the parameter is rather poor, and predictions
depend non-linearly on the parameter (via trigonometric functions). In
the context of present data those issues have been commented on in
ref.~\cite{Gonzalez-Garcia:2014bfa}, based on preliminary simulations
of the statistical properties of the used $\Delta\chi^2$ statistics. Similar
considerations for future experiments can be found in
refs.~\cite{Schwetz:2006md, Blennow:2014sja}.

In this work we extend the results of
ref.~\cite{Gonzalez-Garcia:2014bfa} and study in detail the
distribution of the relevant test statistics by generating large
samples of pseudo data and constructing confidence intervals or
regions with the correct coverage following the Feldman-Cousins
prescription~\cite{Feldman:1997qc}. We study the behavior as a
function of the unknown true values of $\deltaCP, \theta_{23}$, and
the neutrino mass ordering. In refs.~\cite{Schwetz:2006md,
  Blennow:2014sja} the sensitivity to CP violation has been studied,
whereas in this work we concentrate on the related but different
problem of constructing confidence intervals for $\deltaCP$.  In
addition to analyzing present data, we also investigate the behavior
of the test statistics assuming an increased exposure of the T2K
experiment, to be expected in the timescale of several years.  We
attempt to provide an explanation of our numerical results by
considering the non-linear structure of the relevant oscillation
probabilities including parameter degeneracies.

The outline of the paper is as follows. In section~\ref{sec:data} we
briefly describe the data from the T2K and MINOS experiments,
introduce the relevant test statistics, and discuss the statistical
analysis based on the Monte Carlo simulations.  In
section~\ref{sec:probab} we consider the relevant oscillation
probabilities and provide a discussion about why deviations from
Gaussianity can be expected. Our results analyzing present data are
presented in section~\ref{sec:results}. We discuss the distributions
of the 1-dimensional $\Delta\chi^2$ statistics for $\deltaCP$ and
$\theta_{23}$, finding large non-Gaussian behavior, especially for
$\deltaCP$. Then we construct the 2-dimensional regions in the
($\deltaCP,\theta_{23}$) plane and find them to be much closer to the
Gaussian approximation. In section~\ref{sec:future} we investigate how
this situation will change, once more data become available. We study
the sensitivity of T2K by increasing the exposure by roughly a factor
12 compared to the present one including also anti-neutrino data. We
find even in that situation deviations from the Gaussian approximation
remain significant in certain regions of the parameter space. We
summarize and conclude in section~\ref{sec:conclusions}. In the
appendix we show the impact of first data on anti-neutrinos from
T2K~\cite{T2K-EPS15}, which appeared after the completion of this work.

\section{Description of data and statistical analysis}
\label{sec:data}

In this work we use the data from the long-baseline experiments T2K
and MINOS, both from the appearance and disappearance channels,
including also a small anti-neutrino data sample from MINOS, see
Tab.~\ref{tab:data} for details and references.\footnote{Preliminarry
  results from a T2K anti-neutrino run have been released recently at
  EPS HEP 2015~\cite{T2K-EPS15}, consisting of 3 events in the
  appearance channel. We comment on the impact of these data in the
  appendix.} Our code departs from the re-analysis of the data
developed in the context of the NuFit collaboration and used in
ref.~\cite{Gonzalez-Garcia:2014bfa}. For each of the six data samples
shown in Tab.~\ref{tab:data} we perform a spectral fit, where the
numbers of spectral bins are given in the table. Our predictions of
the event spectra $T_i^r(\Theta)$ have been calibrated in order to
reproduce the expected spectra provided by the collaborations. Here
$r$ runs over the six data samples, $i$ labels the energy bins, and
$\Theta$ collectively denotes the oscillation parameters. Each data
set is described by a $\chi^2$ statistics appropriate for Poisson
distributed data:
\begin{align}
  \chi^2_r(\Theta, a_r) &= 2 \sum_i 
    \left[a_r \, T_i^r(\Theta) - O_i^r + O_i^r \log \frac{O_i^r}{a_r \, T_i^r(\Theta)} \right] \,,\\ 
  \chi^2_r(\Theta) &= \min_{a_r}\left[ \chi^2_r(\Theta, a_r) + 
      \left(\frac{1 - a_r}{\sigma^r_{\rm sys}}\right)^2 \right] \,, \label{eq:chisq}
\end{align}
where $O^r_i$ is the observed number of events, and $\sigma^r_{\rm
  sys}$ is the systematic over-all normalization error included via
the pull parameters $a_r$. Since those data are largely statistics
dominated, this simple treatment of systematic errors suffice. For
each experiment our analysis is validated by checking that when
analyzing the data in the same way we can reproduce the confidence
regions in parameter space obtained by the experimental collaborations
with good accuracy.
When combining the data samples given in Tab.~\ref{tab:data} we simply add the $\chi^2$ functions,
\begin{equation}
  \chi^2(\Theta) = \sum_r \chi^2_r(\Theta) \,,
\end{equation}
ignoring possible correlated systematic errors between the data sets.
The results from the reactor experiments~\cite{An:2013zwz, Ahn:2012nd,
  Abe:2014bwa} are taken into account implicitly by fixing
$\theta_{13} = 8.5^\circ$. 

\begin{table}
\centering
  \begin{tabular}{lcrccr}
\hline\hline
Experiment & Channel & Exposure (p.o.t.) & Ref. & Data points & Events \\
\hline
    T2K & $\nu_\mu \to \nu_\mu$ & $6.57\times 10^{20}$ & \cite{Abe:2014ugx} & 16 & 120 \\
    T2K & $\nu_\mu \to \nu_e$   & $6.57\times 10^{20}$ & \cite{Abe:2013hdq} &  5 & 28 \\
    MINOS & $\nu_\mu \to \nu_\mu$ & $10.71\times 10^{20}$ & \cite{Adamson:2013whj} &  39 & 2782 \\
    MINOS & $\overline\nu_\mu \to \overline\nu_\mu$ & $3.36\times 10^{20}$ & \cite{Adamson:2013whj} &  14 & 222\\
    MINOS & $\nu_\mu \to \nu_e$ & $10.6\times 10^{20}$ & \cite{Adamson:2013ue} &  5 & 88 \\
    MINOS & $\overline\nu_\mu \to \overline\nu_e$ & $3.3\times 10^{20}$ & \cite{Adamson:2013ue} &  5 & 9 \\
\hline\hline
  \end{tabular}
  \mycaption{Summary of used data. The last two columns give the
    number of bins used to fit the energy spectrum, and the total
    number of observed events, respectively. Recent T2K data on the
    $\overline\nu_\mu \to \overline\nu_e$ channel \cite{T2K-EPS15} are
    not used in the main text, but we show some results including them
    in the appendix.
\label{tab:data}}
\end{table}

Below we are going to focus on the parameters $\deltaCP$, $\theta_{23}$,
and $\Delta m^2_{32}$, which currently have the largest uncertainties,
including the sign of $\Delta m^2_{32}$. The other oscillation
parameters are fixed to $\theta_{12} = 33.5^\circ$, $\theta_{13} =
8.5^\circ$, $\Delta m^2_{21} = 7.5\times 10^{-5} \, {\rm eV}^2$. Those
parameters are known within better than 15\% at
$3\sigma$~\footnote{Here we define the precision by $2(x^{\rm up} -
  x^{\rm low})/(x^{\rm up} + x^{\rm low})$, where $x^{\rm up}$ and
  $x^{\rm low}$ are the upper and lower ends of the $3\sigma$
  intervals~\cite{Gonzalez-Garcia:2014bfa}, respectively.}  and we
expect that fixing those parameters has only a small impact on our
results. Hence, in the notation used above we have $\Theta = \{\deltaCP,
\theta_{23}, \Delta m^2_{32}\}$.

If we are interested in confidence regions
of one of the parameters $\Theta = \{\deltaCP,
\theta_{23}, \Delta m^2_{32}\}$, irrespective of the others, we
consider the following test statistic. Taking for example $\deltaCP$, we define
\begin{equation}\label{eq:chisq1} 
  \Delta\chi^2(\deltaCP) = \min_{\theta_{23}, \Delta m^2_{32}} \chi^2(\Theta) - 
   \chi^2_{\rm min} \,,
\end{equation}
where $\chi^2_{\rm min}$ is the global minimum of the $\chi^2$ with
respect to all parameters $\Theta$. Note that when minimizing over
$\Delta m^2_{32}$, we always take into account both signs, i.e.\ we
minimize also over the two mass orderings. Similar definitions apply
for the other 1-dimensional cases, $\Delta\chi^2(\theta_{23})$ and
$\Delta\chi^2(\Delta m^2_{32})$. The 2-dimensional confidence regions are
based on an analogous definition, e.g.,
\begin{equation}\label{eq:chisq2} 
  \Delta\chi^2(\deltaCP,\theta_{23}) = \min_{\Delta m^2_{32}} \chi^2(\Theta) - 
   \chi^2_{\rm min} \,.
\end{equation}
This procedure is equivalent to the profile-likelihood method to treat
nuisance parameters.
 
Wilks theorem \cite{wilks} implies that under certain conditions, the
test statistics from eqs.~\eqref{eq:chisq1} and \eqref{eq:chisq2} are
distributed according to the $\chi^2$-distribution with 1 and 2
degrees of freedom (dof), respectively. This is the basis of the
standard method to derive confidence regions for the parameters, using
the condition $\Delta\chi^2 \le t_{\chi^2}({\rm CL, \, dof})$. We
refer to $t_{\chi^2}$ as ``cut levels'', and their values can be
obtained by integrating the corresponding $\chi^2$ distribution. For
instance, the 1-dimensional intervals at 1$\sigma$, 2$\sigma$,
3$\sigma$ are derived by $\Delta\chi^2 \le 1,4,9$,
respectively.\footnote{In this paper we use the two-sided Gaussian
  convention to convert standard deviations into CL, which implies
  that 1$\sigma$, 2$\sigma$, 3$\sigma$ correspond to 68.27\%,
  95.45\%, 99.73\%~CL, respectively.}  We will refer to this situation as
the ``Gaussian limit'' or the ``$\chi^2$ limit'' in the following.

Wilks theorem applies if the theoretical predictions $T_i(\Theta)$
span a linear space when $\Theta$ is varied. For instance, this is
the case if $T_i(\Theta)$ can be expanded to linear order:
$T_i(\Theta) \approx A_i + B_i\Theta$. This is trivially fulfilled for
a linear model, where this relation is exact. For non-linear models
$T_i(\Theta)$, the linear approximation will hold in the vicinity of
the best fit point and will be reliable up to a certain CL, beyond
which the non-linear character of the parameter dependence can lead to
deviations from the Gaussian limit. For ``powerful'' data, which
constrain the parameter efficiently, the linear approximation will
hold up to a high CL, whereas for ``weak'' data with poor sensitivity
to the parameter it will break down already at low CL.  Deviations
from Gaussianity are expected for example close to a physical boundary
of a parameter, or when certain values of the predictions
$T_i(\Theta)$ cannot be reached due to the parameter dependence of the
model, for instance via trigonometric functions, as we are going to
see below.

In order to study deviations from the Gaussian limit we have performed
Monte Carlo (MC) simulations and calculated the distributions of the
test statistics numerically. For assumed true values of the parameters
$\Theta^{\rm true}$ we generate artificial pseudo data by assuming a
Poisson distribution for the observables with mean $T_i^r(\Theta^{\rm
  true})$. Those data are multiplied by a random Gaussian number with
mean 1 and standard deviation $\sigma_{\rm sys}^r$ in order to take
into account the systematic normalization uncertainty. In this
approach the origin of the systematic uncertainty is related to some
auxiliary measurements, determining for instance the fiducial volume
of the detector or the beam normalization. The {\it measured} values
of those experiments are used to determine the theoretical
predictions. Hence, the normalization of the predictions is subject to
statistical fluctuations of the auxiliary measurements. We set the
unknown true value of the normalization constant to 1 and generate
random realisations of this number with standard deviation
$\sigma_{\rm sys}^r$, which then enters the ``observables'' for the MC
generated data.  This implies that we consider the auxiliary
measurements as part of the experiment, also hypothetically to be
repeated many times.  However, we have checked that our results do not
depend on whether we randomize the systematic error or not.

Then the test statistics eqs.~\eqref{eq:chisq1} or \eqref{eq:chisq2}
are calculated at the point $\Theta^{\rm true}$, e.g., $\Delta\chi^2_{\rm
  MC[\Theta]}(\deltaCP)$, where the subscript MC[$\Theta$] indicates
that the pseudo data from the Monte Carlo generated at the point
$\Theta$ are used in the $\chi^2$. We perform this calculation $10^4$
times for each point in the $\Theta^{\rm true}$ space, which provides
us the true distribution of the test statistic for each parameter
value. From those histograms we can obtain the true cut levels $t_{\rm
  MC}({\rm CL}, \Theta)$ for a given CL $\alpha$, by demanding that a
fraction $\alpha$ of all pseudo experiments fulfills
\begin{equation}\label{eq:tMC}
\Delta\chi^2_{\rm MC[\Theta]}(\deltaCP) \le t_{\rm MC}({\rm CL}, \Theta) \, .   
\end{equation}
Then the correct confidence intervals (or regions) for the parameters are
obtained by those values of the parameters for which the test
statistics of the real data fulfills $\Delta\chi^2(\deltaCP) \le
t_{\rm MC}({\rm CL}, \Theta)$. Those intervals have the correct
coverage by construction and follow the prescription of Feldman and
Cousins~\cite{Feldman:1997qc}.  

Here we used the test statistic for
$\deltaCP$ as an example, but analogous expressions hold for the
other 1-dimensional as well as 2-dimensional test statistics. Note
that in the left side of eq.~\eqref{eq:tMC}, $\Delta\chi^2$ is
evaluated at $\deltaCP$ corresponding to the same value as used to
generate the pseudo data. However, although the test statistic for
given data depends only on $\deltaCP$, the MC results do depend also
on the other parameters $\theta_{23}$ and $\Delta m^2_{32}$. Hence,
the confidence intervals for $\deltaCP$ may depend on the unknown true
values of the other parameters, an effect we will indeed observe in
our numerical studies presented in the next section.

\section{Discussion of oscillation probabilities}
\label{sec:probab}

In the case of interest, we are facing a complicated parameter
dependence of the predictions. We review here the relevant oscillation
probabilities through which the parameters enter the event rate
predictions. Note that we do not use the approximate expression for
the numerical work (which is based on numerical calculations of the
full three-flavor probabilities including the matter effect) but they
serve well for a qualitative understanding.

Let us define 
\begin{equation}
\Delta \equiv \frac{|\Delta m^2_{31}| L}{4E} \,,\quad
A \equiv \left| \frac{2EV}{\Delta m^2_{31}} \right|\,,
\end{equation}
where $L$ is the baseline, $E$ is the neutrino energy, and $V$ is the
effective matter potential. For the $\nu_\mu$ disappearance channel we have
\begin{align}
  P_{\rm dis} \approx& \sin^22\theta_{23} \, \sin^2\Delta \,, \label{eq:Pdiss}
\end{align}
where we show only the leading term and neglect
corrections due to $\Delta_{21}$ as well as $\theta_{13}$.
An approximate expression for the $\nu_\mu\to\nu_e$ oscillation
probability, valid for a constant matter
density is given by~\cite{Cervera:2000kp,Freund:2001pn}:
\begin{align}
P_{\rm app} &\approx 4 \, \sin^2\theta_{13} \, \sin^2\theta_{23} \frac{\sin^2
         \Delta (1 - asA)}{(1 - asA)^2}  \nonumber\\
         &+  s\, \frac{\Delta m^2_{21}}{|\Delta m^2_{31}|} \, \sin2\theta_{13} \, \sin 2\theta_{12} \, 
         \sin2\theta_{23} \cos(s\Delta + a \deltaCP) \, 
	 \frac{\sin\Delta A}{A} \, \frac{\sin \Delta (1 - asA)}{1-asA} \,.
\label{eq:P}
\end{align}
The signs
$a$ and $s$ describe the effects of CP-conjugation and the
neutrino mass ordering, respectively, with
$a = +1$ for neutrinos and $a=-1$ for anti-neutrinos, and 
$s = \mathrm{sgn}(\Delta m^2_{31})$. 
The matter effect enters via the parameter $A$.  
Numerically one finds for a matter
density of 3~g/cm$^3$
\begin{equation}\label{eq:A}
A \simeq 0.094 \, 
\left(\frac{E}{\rm GeV}\right)
\left(\frac{|\Delta m^2_{31}|}{2.4\times 10^{-3} 
\: \mathrm{eV}^2}\right)^{-1} \,.
\end{equation} 
Hence, for the T2K experiment, with $E \simeq 0.7$~GeV, the matter
effect is of order 6\% and we can expand eq.~\eqref{eq:P} also in
$A$, keeping only terms up to first order in $A$. To simplify the
expression further we assume the first oscillation maximum, $\Delta
\approx \pi/2$, which is a good approximation for
T2K. Introducing the definitions
\begin{equation}
  \sin^2\theta_{23} = \frac{1}{2} + d \,,\qquad
  C \equiv \frac{\Delta m^2_{21} L}{4E} \, \sin2\theta_{13} \sin 2 \theta_{12} \sin 2 \theta_{23} \,, 
\end{equation}
eq.~\eqref{eq:P} becomes for neutrinos and anti-neutrinos
\begin{align}
  P_\nu      &\approx 2\sin^2\theta_{13}(1 + 2d)(1+2sA) - C \,\sin\deltaCP \,(1+sA) \,,
  \label{eq:Pn}\\
  P_{\bar\nu} &\approx 2\sin^2\theta_{13}(1 + 2d)(1-2sA) + C \,\sin\deltaCP \,(1-sA) \,. 
  \label{eq:Pa}
\end{align}

Note that the magnitude of $d$ is constrained by the
$\nu_\mu\to\nu_\mu$ disappearance channel, but we are left with the
octant degeneracy for $\theta_{23}$, described by a sign ambiguity of
$d$, with $d < 0$ ($d > 0$) corresponding to the first (second) octant
for $\theta_{23}$. We are going to consider values in the range $-0.1
\le d \le 0.1$, within the currently $3\sigma$ confidence interval. Hence,
the free parameters in the problem are the continuous parameter
$\deltaCP$ and the two signs of $d$ and $s$, i.e.\ four discrete sign
combinations. Note, however, that especially for current data this is
an over-simplification, since the uncertainty on $|d|$ is large.
Numerically we have $\sin^2\theta_{13} \approx 0.022$, $C \approx
0.013$ (with a very weak dependence on $d$ for $|d| \lesssim 0.1$),
and $A \approx 0.06$. Hence, all terms in eqs.~\eqref{eq:Pn} and
\eqref{eq:Pa} are of similar order, and both the
octant~\cite{Fogli:1996pv} and the mass
ordering~\cite{Minakata:2001qm} degeneracies will lead to changes in
the predictions of similar size as $\deltaCP$. Recent discussions of
the $\deltaCP$ and $\theta_{23}$ interplay can be found in
refs.~\cite{Minakata:2013eoa, Coloma:2014kca, Ghosh:2015ena}.

Given the parameter dependencies from eqs.~\eqref{eq:Pdiss} and (\ref{eq:Pn}, \ref{eq:Pa})
we can expect deviations from the Gaussian limit for the following reasons.
\begin{enumerate}
\item 
Present data show only weak sensitivity to $\deltaCP$, i.e., the full
range $0 \le \deltaCP < 2\pi$ is allowed at relatively low CL. This
implies that the strong non-linearity of the trigonometric dependence
in eq.~\eqref{eq:Pn} comes into play. In particular, because of the
sine dependence, only a finite change of $T_i(\deltaCP)$ can be
achieved by varying $\deltaCP$, in contrast to the unbounded variation of a linear
model. This means that $\deltaCP$ effectively provides less than
1~dof, which implies that the true cut levels for a given CL will be
lower than the ones corresponding to the Gaussian approximation.

For future data with decreased statistical errors the fluctuations of
the data may become of similar size as the compact region spanned in
$T_i$ upon varying $\deltaCP$, which means that the coverage of $T_i$
is actually more efficient than for the linear model and $\deltaCP$
thus provides more than 1~dof. If the exposure is further increased the
fluctuations will become even smaller than the compact region in $T_i$
such that the linear expansion becomes valid and we approach 1~dof.

\item
Disappearance data depend on $\sin^22\theta_{23}$, which leads to the
octant degeneracy for $\theta_{23}$. However, due to the
$\sin^2\theta_{23}$ factor in the appearance probability the
degeneracy is not complete when $\nu_\mu\to\nu_e$ data are
included. The two possible solutions for $\sin^2\theta_{23}$
corresponding to the two signs of $d$ imply more freedom than a single linear
parameter. Hence the presence of the degeneracy leads to an increase
of the effective dofs, which implies increased cut levels.

\item
For values of $\theta_{23}$ close to $\pi/4$ we are facing a physical
boundary in the disappearance channel, since $\sin^22\theta_{23} \le
1$, which implies that predictions corresponding to
$\sin^22\theta_{23} > 1$ cannot be reached by varying
$\theta_{23}$. This leads to a decrease of the effective dof, and
reduces the cut levels.

\item
Similarly, for $\deltaCP \simeq \pi/2$ or $3\pi/2$ the $\sin\deltaCP$
dependence in the appearance probabilities \eqref{eq:Pn} and
\eqref{eq:Pa} imply a physical boundary due to $|\sin\deltaCP| \le 1$.
For those values of $\deltaCP$ the derivatives of the probabilities
with respect to $\deltaCP$ vanish. Hence, we expect decreased cut
levels for those values, while for $\deltaCP \simeq 0$ or $\pi$ the
dependence of the appearance channel resembles approximately a linear
model. The behavior around $\deltaCP \simeq \pi/2$ or $3\pi/2$ is
further complicated by the octant and mass ordering degeneracies, see
the discussion related to eq.~\eqref{eq:Pminmax} below.
\end{enumerate}

In our numerical results presented below we will observe all of those
effects, where some of them may occur simultaneously, leading to a
complicated interplay of effects. Nevertheless, some general features
can be understood qualitatively. For instance, considering the
$\deltaCP$ dependence of eqs.~\eqref{eq:Pn} and \eqref{eq:Pa} plus the
4-fold degeneracy related to the signs of $d$ and $s$, we find that
there are minimal and maximal values for the oscillation probabilities
(and hence for the event rates) given by the following combinations of
the parameters:
\begin{align}
  \begin{array}{llll}    
  P_\nu^{\rm max}:\quad &  d > 0 \, (\text{2nd oct.})\,,\quad &s = +1 \,(\text{NO})\,, \quad &\deltaCP = 3\pi/2 \\ 
  P_\nu^{\rm min}:\quad  &  d < 0 \,(\text{1st oct.})\,,\quad &s = -1 \,(\text{IO})\,, \quad &\deltaCP = \pi/2 \\[2mm] 
  P_{\bar\nu}^{\rm max}:\quad  &  d > 0 \,(\text{2nd oct.})\,,\quad &s = -1 \,(\text{IO})\,, \quad &\deltaCP = \pi/2\\ 
  P_{\bar\nu}^{\rm min}:\quad  &  d < 0 \,(\text{1st oct.})\,,\quad &s = +1 \,(\text{NO})\,, \quad &\deltaCP = 3\pi/2 
  \end{array}\label{eq:Pminmax}  
\end{align}
If the true values of $d,s,\deltaCP$ correspond to one of the
combinations in eqs.~\eqref{eq:Pminmax} then we are located at a
physical boundary for the event rates: there is no point in the
parameter space which can provide a larger (or smaller) value of the
probability. Statistical fluctuations leading to even larger (or
smaller) event rates than predicted for those extreme parameter values
cannot be accommodated by adjusting the model.  This implies that the
effective number of dof of the $\Delta\chi^2$ is reduced,
i.e.\ lower cut levels. A related discussion can also be found in
ref.~\cite{Blennow:2014sja}.

\section{Results for present data}
\label{sec:results} 

\subsection{One-dimensional intervals for $\deltaCP$}
\label{sec:1d}

\begin{figure}[t!]
 \centering
 \includegraphics[width=\textwidth]{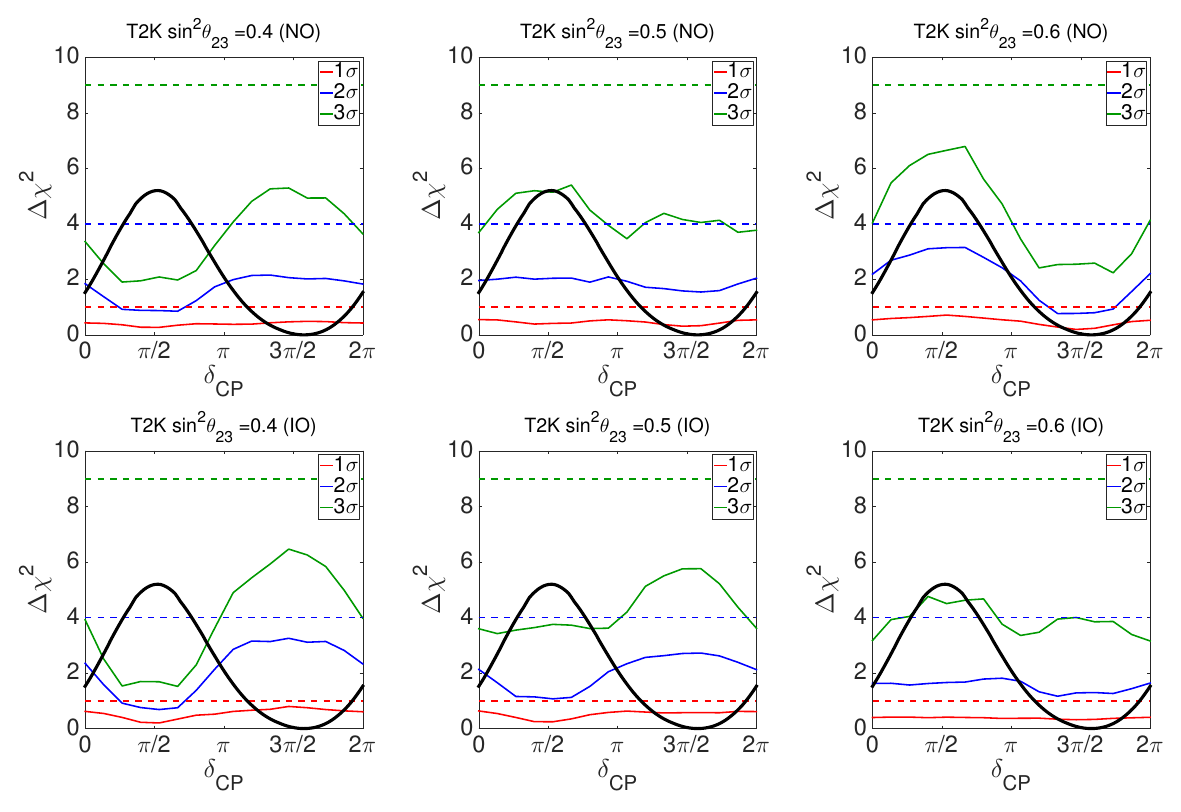}
  \mycaption{The cut levels $t_{\rm MC}({\rm CL}, \Theta)$
    for $\Delta\chi^2(\deltaCP)$ from T2K data for $1\sigma$ (red),
    $2\sigma$ (blue), $3\sigma$ (green). Dashed lines indicate the
    Gaussian approximation $t_{\chi^2}$. Left, middle, right panels
    correspond to $\sin^2\theta_{23}^{\rm true} = 0.4, 0.5, 0.6$,
    respectively. We take $|{\Delta m^2_{32}}^{\rm true}| = 2.4\times
    10^{-3} \, {\rm eV}^2$ and for the upper (lower) row we have
    assumed a true normal (inverted) mass ordering. The black solid
    curve shows $\Delta\chi^2(\deltaCP)$ using the observed data (same
    curve in all panels).
\label{fig:1d-T2K} 
} 
\end{figure}

\begin{figure}[t!]
 \centering
 \includegraphics[width=\textwidth]{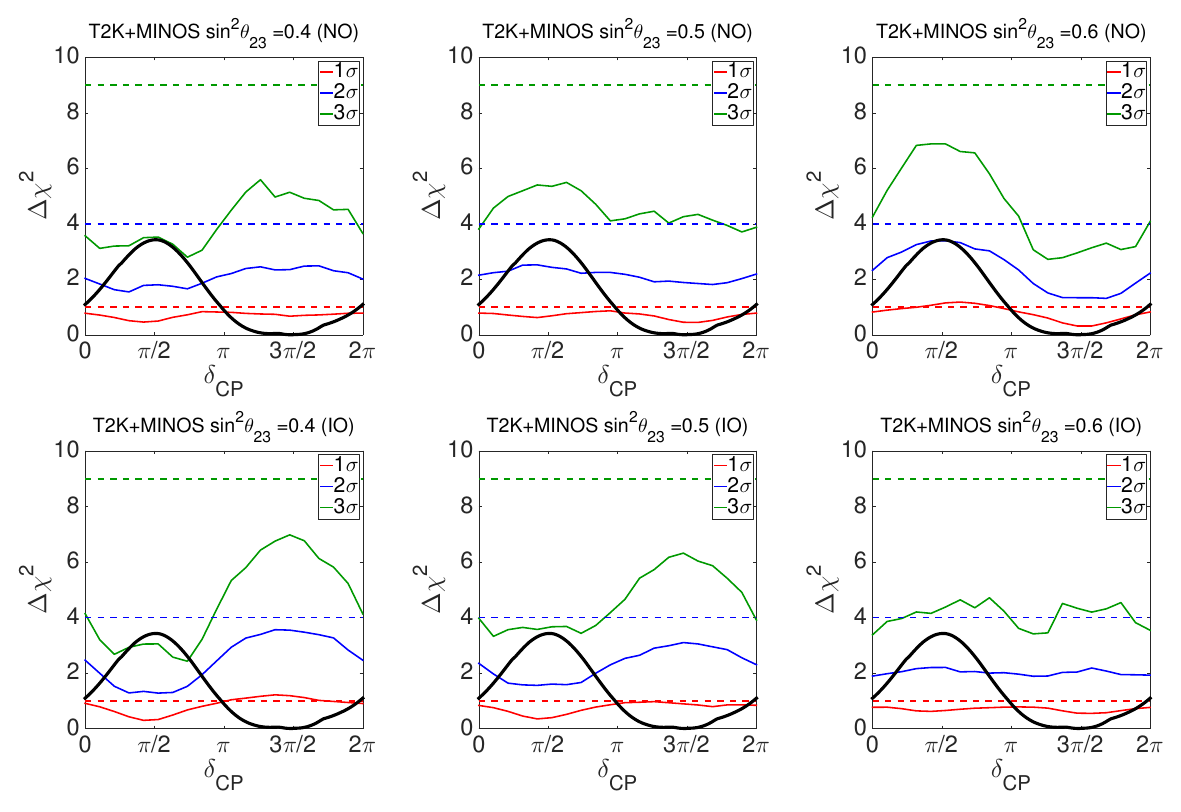}
  \mycaption{Same as fig.~\ref{fig:1d-T2K} but using combined T2K and MINOS data.
   \label{fig:1d-T2K+MINOS}}
\end{figure}

We start presenting the results of our simulations for the
1-dimensional $\Delta\chi^2$ distributions defined in
eq.~\eqref{eq:chisq1}. In figs.~\ref{fig:1d-T2K} and
\ref{fig:1d-T2K+MINOS} we consider $\Delta\chi^2(\deltaCP)$ for the CP
phase $\deltaCP$ and show the cut levels $t_{\rm MC}$ for $1\sigma$,
$2\sigma$, $3\sigma$. Fig.~\ref{fig:1d-T2K} uses T2K data only whereas
fig.~\ref{fig:1d-T2K+MINOS} uses all data given in
Tab.~\ref{tab:data}, showing qualitatively similar results to the
T2K-only case.  We have checked that our T2K results are consistent
with the ones shown by the T2K collaboration~\cite{Abe:2015awa}, in
cases where comparison is possible.

By comparing the curves for $t_{\rm MC}$ to the $\chi^2$ approximation
$t_{\chi^2}$ indicated by the dashed curves, we observe significant
deviations for the Gaussian limit, with $t_{\rm MC}(2\sigma)$ and
$t_{\rm MC}(3\sigma)$ being much lower than the corresponding
$t_{\chi^2}$~\cite{Gonzalez-Garcia:2014bfa}. Furthermore we find large
variations of the $\Delta\chi^2(\deltaCP)$ distribution depending on
$\deltaCP$ itself, and on the assumed true values for
$\sin^2\theta_{23}$~\cite{Gonzalez-Garcia:2014bfa} and to a lesser
extent also depending on the mass ordering. The various panels in
figs.~\ref{fig:1d-T2K} and \ref{fig:1d-T2K+MINOS} correspond to
different assumptions about $\theta_{23}^{\rm true}$ and the true mass
ordering. Note that those are the ``true'' values assumed for
generating the pseudo data, while when fitting to the data we leave
$\theta_{23}$ and $\Delta m^2_{32}$ free. Comparing
figs.~\ref{fig:1d-T2K} and \ref{fig:1d-T2K+MINOS}, we find that the
addition of MINOS data makes some of the ``dips'' in the cut levels
less sever, e.g.\ the ones for $\sin^2\theta_{23} \simeq 0.4$ and
$\delta \simeq \pi/2$ for both orderings.

The behavior of the $t_{\rm MC}$ curves can be understood from the
discussion given in section~\ref{sec:probab}. The reduction of
$t_{\rm MC}$ compared to the Gaussian limit follows from the
poor sensitivity of the data to $\deltaCP$, which implies that the full
range $0\le\deltaCP < 2\pi$ becomes accessible. Hence the
trigonometric dependence becomes relevant, changing
$\deltaCP$ provides less freedom than a linear parameter, and the
effective number of dof becomes reduced. 

\begin{figure}[t!]
\includegraphics[width=\textwidth]{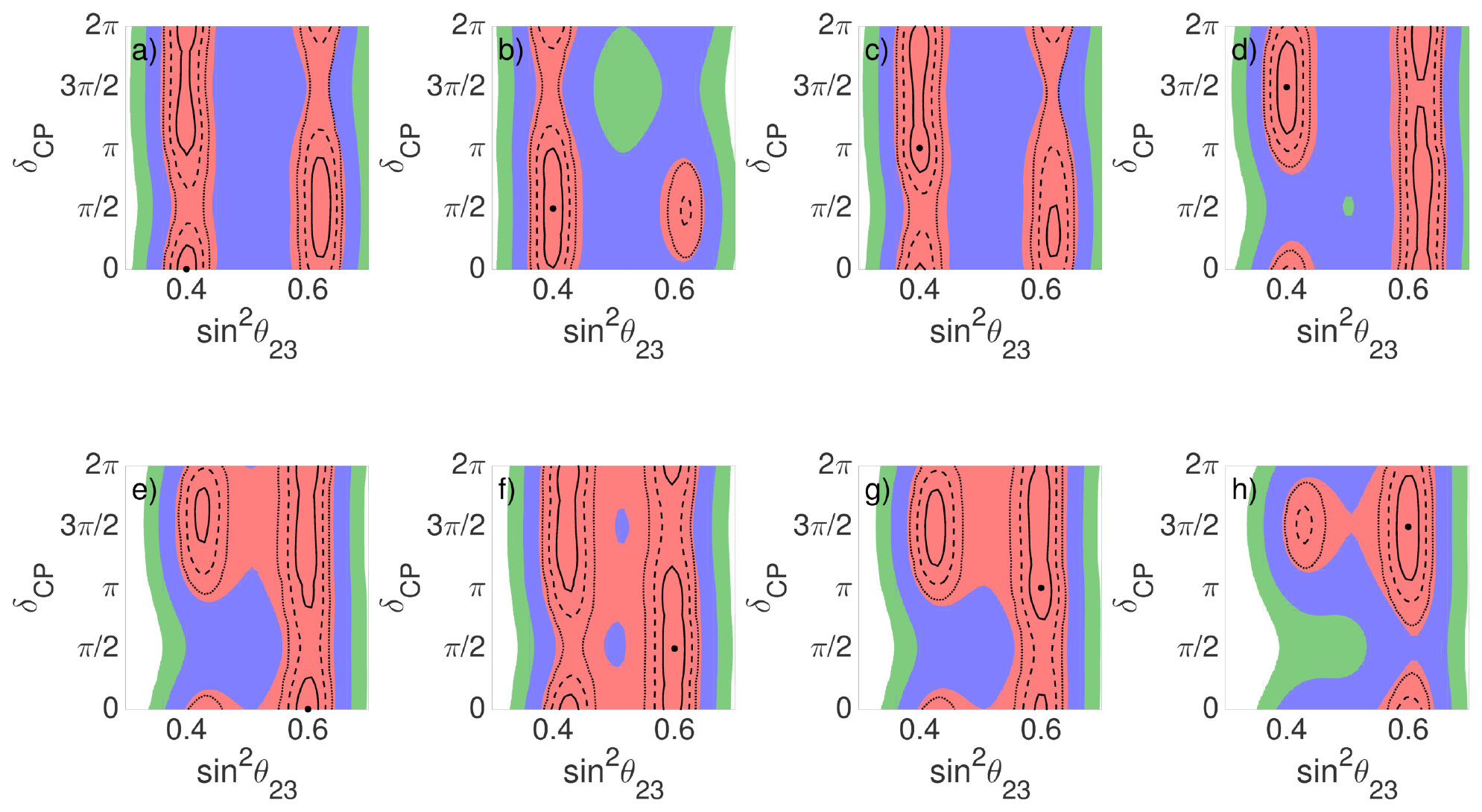}
 \mycaption{Sensitivity of T2K data in the ($\sin^2\theta_{23}$,
   $\deltaCP$) plane based on Asimov data at $1\sigma$, $2\sigma$,
   $3\sigma$ assuming $\sin^2\theta_{23}^{\rm true} = 0.4$ (upper row)
   and 0.6 (lower row) for different values of $\deltaCP^{\rm true}$
   as marked by the dots in the plots. We assume ${\Delta
     m^2_{32}}^{\rm true} = 2.4\times 10^{-3} \, {\rm eV}^2$
   (NO). Colored areas correspond to the current exposure, whereas the
   black contour curves correspond to an exposure of
   7.8$\times10^{21}$ p.o.t.\ neutrino data (about 12 times the
   current exposure). Regions are derived by assuming the Gaussian
   approximation.
\label{fig:sensitivity}}
\end{figure}

The appearance of the bumps, for instance for normal ordering and
($\sin^2\theta_{23}^{\rm true} = 0.4, \deltaCP \simeq 3\pi/2$) and
($\sin^2\theta_{23}^{\rm true} = 0.6, \deltaCP \simeq \pi/2$) can be
understood by considering the $\theta_{23}$ octant degeneracy. We show
in fig.~\ref{fig:sensitivity} the sensitivity of T2K in the
($\sin^2\theta_{23}$, $\deltaCP$) plane.  We calculate so-called Asimov
data, using the theoretical prediction for certain assumed true values
without statistical fluctuations as ``data''. This indicates the
expected sensitivity for that particular set of true values for an average
experiment. Regions in fig.~\ref{fig:sensitivity} are derived based on
the Gaussian approximation for $\Delta\chi^2(\deltaCP, \theta_{23})$. 

By comparing those results with the upper row of panels in
fig.~\ref{fig:1d-T2K} we find a correlation with the ability to
resolve the degeneracy: in cases with improved sensitivity to the
octant (panels (b) and (h) in fig.~\ref{fig:sensitivity}) the cut
levels are low (upper left panel for $\deltaCP \simeq \pi/2$ and upper
right panel for $\deltaCP \simeq 3\pi/2$ in fig.~\ref{fig:1d-T2K}),
whereas for cases where the degeneracy is strong for all values of
$\deltaCP$ (panels (d) and (f) in fig.~\ref{fig:sensitivity}) the cut
levels are high (upper left panel for $\deltaCP \simeq 3\pi/2$ and
upper right panel for $\deltaCP \simeq \pi/2$ in
fig.~\ref{fig:1d-T2K}). This shows that the presence of the degeneracy
increases the effective number of dof for the $\Delta\chi^2$
distribution.  We have confirmed a similar correspondence also for the
inverted ordering. Note that the overall sensitivity to the octant is
very poor even in the cases shown in panels (d) and (f) in
fig.~\ref{fig:sensitivity}, corresponding to $\Delta\chi^2 \approx
0.5$ for the wrong octant solution. However, for the distribution of
$\Delta\chi^2(\deltaCP)$ it is relevant that the degeneracy can be
resolved for a significant range of $\deltaCP$ values, while for the
the cases corresponding to panels (d) and (f) in
fig.~\ref{fig:sensitivity} the degeneracy is present at below
$1\sigma$ for all values of $\deltaCP$.  Those results can also be
understood from the discussion related to the maximal and minimal
values of the oscillation probability given in the first 2 lines of
eq.~\eqref{eq:Pminmax}. Fig.~\ref{fig:1d-T2K} shows low cut levels for
those combinations of parameters, following from the physical boundary
of the event rates.\footnote{The reduced cut levels for NO,
  $\sin^2\theta_{23}=0.4$, $\deltaCP \simeq \pi/2$ seen in the left
  upper panel of fig.~\ref{fig:1d-T2K} do not follow from this
  argument. While $d < 0$ and $\deltaCP \simeq \pi/2$ does correspond
  to a minimum of the oscillation probability, apparently in that case
  changing the mass ordering from NO to IH does not provide enough
  freedom to overcome the physical boundary.}

For the cut levels in figs.~\ref{fig:1d-T2K} and
\ref{fig:1d-T2K+MINOS}, we have always minimized with respect to both
mass orderings. However, we have checked that assuming that the mass
ordering was known (i.e., restricting to the true ordering in the fit)
does not change the results qualitatively. This is different from the
cut levels for increased exposure discussed in
section~\ref{sec:future}, where we will find a significant effect of
restricting the mass ordering.

\begin{table}[t]
\centering
\begin{tabular}{ccccc}  
\hline\hline
true $\sin^2\theta_{23}$ & true MO & $1\sigma$& $2\sigma$& $3\sigma$ \\
\hline
0.4 & normal & [3.20, 6.00] & [0, 0.42] $\cup$ [2.65, 2$\pi$] & [0, 2$\pi$] \\
0.5 & normal & [3.21, 5.94] & [0, 0.70] $\cup$ [2.48, 2$\pi$] & [0, 2$\pi$] \\
0.6 & normal & [3.16, 5.90] & [0, 2$\pi$] & [0, 2$\pi$] \\
0.4 & inverted & [3.11, 6.14] & [0, 0.44] $\cup$ [2.63, 2$\pi$] & [0, 1.08] $\cup$ [2.35, 2$\pi$] \\
0.5 & inverted & [3.15, 6.07] & [0, 0.45] $\cup$ [2.61, 2$\pi$] & [0, 2$\pi$] \\
0.6 & inverted & [3.23, 5.92] & [0, 0.55] $\cup$ [2.59, 2$\pi$] & [0, 2$\pi$] \\
\hline
\multicolumn{2}{c}{Gaussian limit} & [3.09, 6.20] & [0, 2$\pi$] & [0, 2$\pi$] \\
\hline\hline
\end{tabular}
  \mycaption{Confidence intervals at $1\sigma$, $2\sigma$, $3\sigma$
    for the CP phase $\deltaCP$ from combined T2K and MINOS data for
    different assumptions about the true values of $\theta_{23}$ and
    the neutrino mass ordering. The last row shows the confidence
    intervals in the Gaussian approximation.
   \label{tab:deltaCIs}}
\end{table}

\begin{figure}[t]
 \centering
 \includegraphics[width=0.7\textwidth]{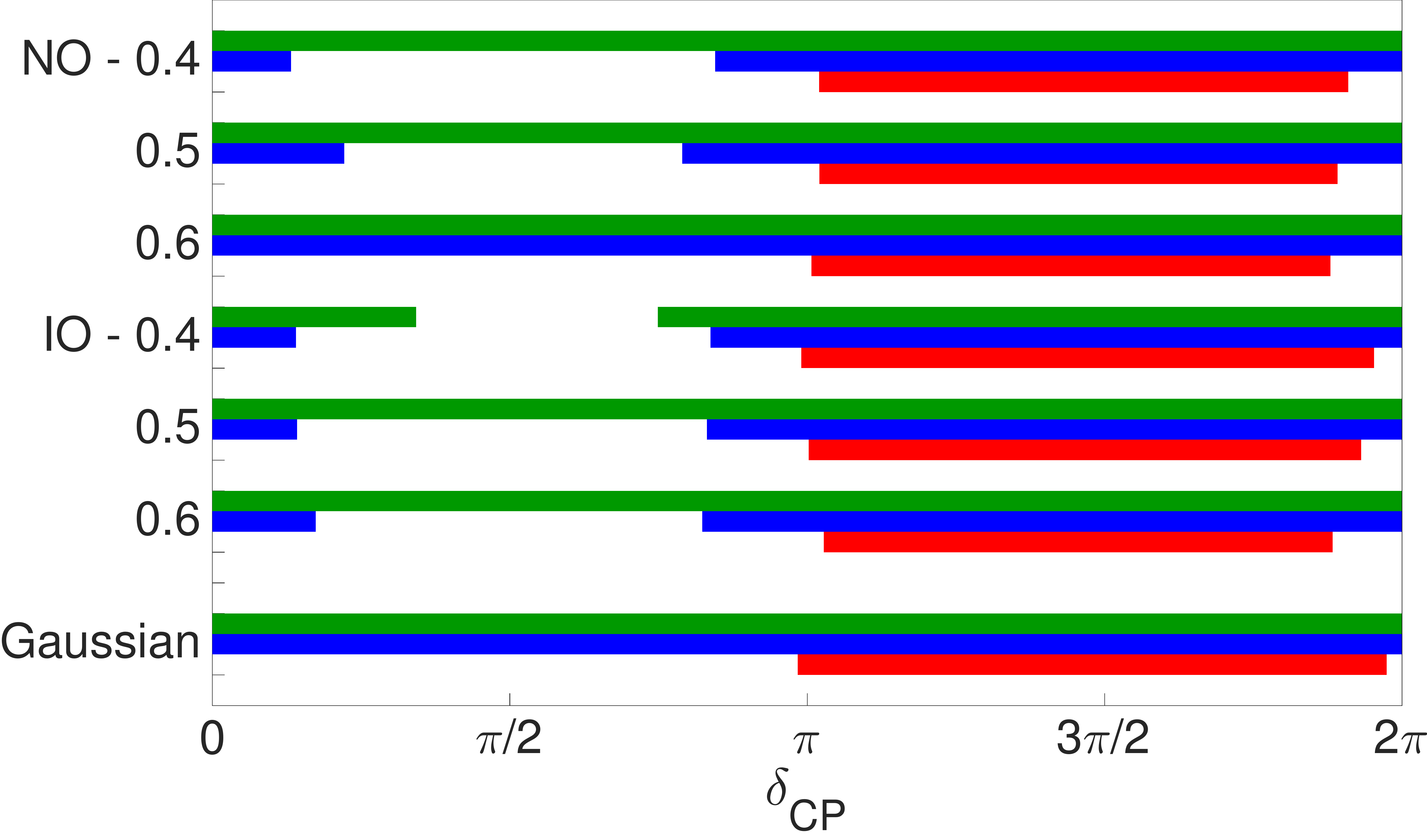}
 \mycaption{Graphical representation of the confidence intervals for
   $\deltaCP$ at $1\sigma$ (red), $2\sigma$ (blue), $3\sigma$ (green),
   see table~\ref{tab:deltaCIs}. The labels on the left side indicate
   the true value of $\sin^2\theta_{23}$ and the mass ordering. The
   bottom three bars correspond to the Gaussian approximation.
   \label{fig:deltaCIs}}
\end{figure}

The exact confidence intervals for $\deltaCP$ can be obtained by
comparing the $\Delta\chi^2(\deltaCP)$ from the observed data (shown
as black curves in figs.~\ref{fig:1d-T2K} and \ref{fig:1d-T2K+MINOS})
to the curves for the cut levels. The confidence interval at a given
CL is obtained by those values of $\deltaCP$ for which
$\Delta\chi^2(\deltaCP) \le t_{\rm MC}({\rm CL})$.  We show the
confidence intervals for $\deltaCP$ at $1\sigma$, $2\sigma$, $3\sigma$
in table~\ref{tab:deltaCIs} and fig.~\ref{fig:deltaCIs} for different
assumptions about the true values of $\theta_{23}$ and the mass
ordering, and we compare them to the Gaussian approximation.  Because
of the dependence of $t_{\rm MC}$ on the values of the other
parameters the confidence intervals for $\deltaCP$ depend on those
unknown true values. From fig.~\ref{fig:deltaCIs} we see that the
$1\sigma$ intervals are relatively stable and agree well with the
Gaussian approximation. The variations are relatively large for the
$2\sigma$ interval. There is a large parameter dependence on the CL of
rejection of $\deltaCP\simeq \pi/2$. Using combined T2K+MINOS data
this rejection ranges from around $3\sigma$ for
$\sin^2\theta_{23}=0.4$ (both mass orderings) to only $2\sigma$ for
$\sin^2\theta_{23}=0.6$ (NO), see fig.~\ref{fig:1d-T2K+MINOS}. We
obtain $\Delta\chi^2 = 3.37$ at $\deltaCP = \pi/2$. Hence, the
Gaussian approximation gives a rejection of $\deltaCP = \pi/2$ at
$1.8\sigma$. The dependence of the MC results on the true values of
other parameters is even more pronounced for T2K data only, see
fig.~\ref{fig:1d-T2K}. Note that the rejection of $\deltaCP \simeq
\pi/2$ is stronger for T2K data-only and the significance decreases
somewhat when MINOS data are included. The reason for the slight
decrease of $\Delta\chi^2(\deltaCP)$ when MINOS data are added to T2K
is that MINOS appearance data prefer a somewhat smaller value of
$\theta_{13}$ than T2K and hence the combination with the reactor
result for $\theta_{13}$ becomes somewhat less effective in
constraining $\deltaCP$ when MINOS is included.

Within the frequentist framework there is no way to marginalize or
average over the true values of nuisance parameters, since those are
considered to be fixed constants of Nature. Hence the dependence of
the results for $\deltaCP$ on the unknown true values of other
parameters (especially $\theta_{23}$) introduces an unpleasant
ambiguity. One possibility to deal with this situation could be to
present for each CL the largest confidence interval for
$\deltaCP$. Then the CL would be a lower bound on the true coverage of
the interval, i.e., such an interval at the $\alpha$ CL will cover the
true value with a probability of at least $\alpha$. A problem with
this approach is that one has to maximize the confidence interval of
$\deltaCP$ with respect to the true value of $\theta_{23}$, and within
a frequentist framework it is not possible to decide which range for
$\theta_{23}$ has to be considered.\footnote{A similar discussion in
  the context of the mass ordering determination can be found in
  ref.~\cite{Blennow:2013oma}.}  This problem is solved by considering
2-dimensional confidence regions in both parameters. Before presenting
those in section~\ref{sec:2dt} below, we proceed now by discussing the
1-dimensional intervals for $\theta_{23}$.

\subsection{One-dimensional intervals for $\theta_{23}$}
\label{sec:1th}

\begin{figure}[t!]
\includegraphics[width=\textwidth]{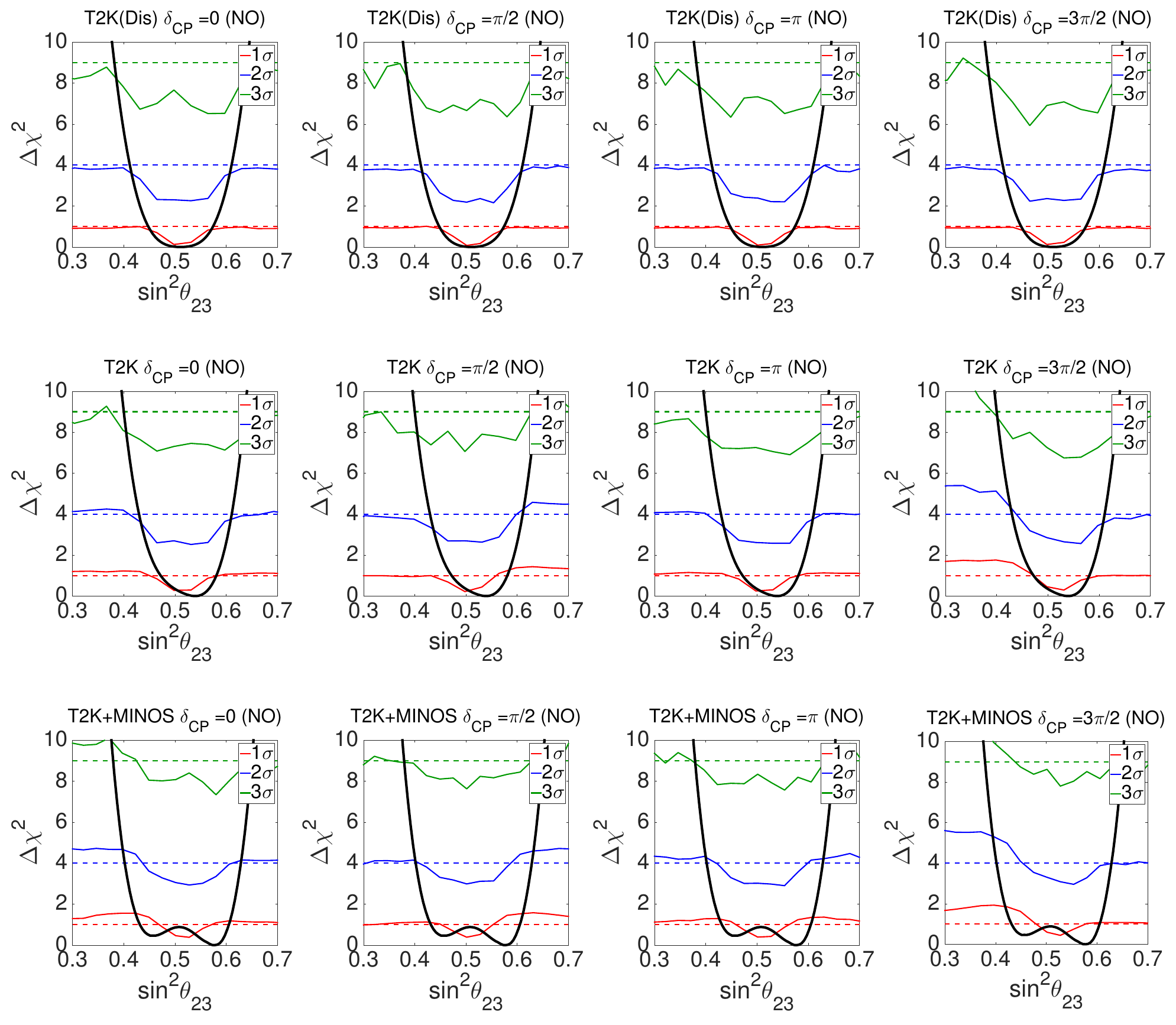}  
  \mycaption{The cut levels $t_{\rm MC}({\rm CL}, \Theta)$
    for $\Delta\chi^2(\theta_{23})$ for $1\sigma$ (red), $2\sigma$
    (blue), $3\sigma$ (green). Dashed lines indicate the Gaussian
    approximation $t_{\chi^2}$. Top row: T2K disappearance only,
    middle row: T2K disappearance and appearance, bottom row: combined
    T2K and MINOS data. The columns correspond to $\deltaCP^{\rm true} =
    0, \pi/2, \pi, 3\pi/2$ from left to right, and we always assume a
    true normal mass ordering with ${\Delta m^2_{32}} =
    2.4\times 10^{-3} \, {\rm eV}^2$. The black solid curve shows
    $\Delta\chi^2(\theta_{23})$ using the observed data (same curve in
    the 4 panels in each row).
\label{fig:1theta}}
\end{figure}

The cut levels for $\Delta\chi^2(\theta_{23})$ obtained from our MC
simulation are shown in fig.~\ref{fig:1theta} for T2K disappearance
data only (top row), for T2K disappearance and appearance data (middle
row), and T2K and MINOS combined (bottom row). The columns of panels
correspond to different values of the CP phase $\deltaCP$ used to
generate the pseudo data. We make the following observations:
\begin{enumerate}
\item For non-maximal values $|\sin^2\theta_{23} - 0.5| \gtrsim 0.05$,
  the test statistic $\Delta\chi^2(\theta_{23})$ is distributed
  approximately as a $\chi^2$ with 1~dof, according to the Gaussian
  approximation.
\item For maximal values $\sin^2\theta_{23} \simeq 0.5$ the cut levels
  are somewhat reduced compared to the Gaussian case. This is the
  manifestation of the physical boundary $\sin^22\theta_{23} \le 1$,
  which reduces the freedom provided by the parameter $\theta_{23}$.
\item Disappearance data only (top row) is insensitive to the true value of the phase $\deltaCP$. 
\item When appearance data are included we observe a slight increase
  of the cut levels compared to the Gaussian limit for certain
  combinations of $\theta_{23}$ and $\deltaCP^{\rm true}$, namely for
  $\deltaCP^{\rm true} = \pi/2$, $\sin^2\theta_{23} \gtrsim 0.6$ and
  for $\deltaCP^{\rm true} = 3\pi/2$, $\sin^2\theta_{23} \lesssim
  0.4$. These are the same regions where we have noted also an increase in
  the cut levels for $\Delta\chi^2(\deltaCP)$ in the previous section,
  corresponding to the cases of strong octant degeneracy, see panels
  (d) and (f) in fig.~\ref{fig:sensitivity}.
\item
  The behavior of the cut levels is basically unchanged by adding
  MINOS data to T2K. However, the $\Delta\chi^2(\sin^2\theta_{23})$
  from the observed data (black curves in the plots) is slightly
  disfavoring maximal mixing. While the confidence intervals for
  $\sin^2\theta_{23}$ based on the MC from T2K happen to be very
  similar to the $\chi^2$ approximation, some deviations from the
  Gaussian limit occur when MINOS data are added, since the observed
  $\Delta\chi^2$ is pushed somewhat into regions where $t_{\rm MC}$
  differs from $t_{\chi^2}$.
\end{enumerate}

Fig.~\ref{fig:1theta} corresponds to assuming a normal ordering for
generating the MC data. Very similar results are obtained also for
inverted ordering. We have also investigated the distribution of the
1-dimensional $\Delta\chi^2$ for $\Delta m^2_{23}$ and we have found
very good agreement with the Gaussian limit, independent of any other
parameters for any combination of T2K and/or MINOS data. This reflects
the very robust determination of $\Delta m^2_{32}$ by the $\nu_\mu$
disappearance spectral data.

\subsection{Two-dimensional confidence regions}
\label{sec:2dt}

As we have seen above, the distribution of the test statistic for
$\deltaCP$ depends significantly on the unknown true value of
$\theta_{23}$. Hence, treating $\theta_{23}$ as nuisance parameter
leads to the unpleasant result that the confidence intervals for
$\deltaCP$ cannot be stated independently of the true value of
$\theta_{23}$. One way to deal with such a situation in a frequentist
framework is to consider two-dimensional confidence regions of both
parameters, keeping in mind that the interpretation of the results is
different.

\begin{figure}[t!]
\includegraphics[width=\textwidth]{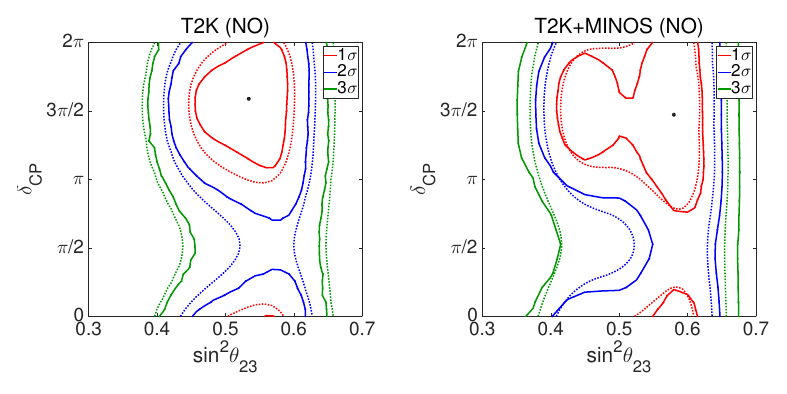}
 \mycaption{Two-dimensional confidence regions at $1\sigma$ (red),
   $2\sigma$ (blue), $3\sigma$ (green) in the $(\sin^2\theta_{23}, \deltaCP)$
   plane for T2K (left panel) and T2K + MINOS (right panel). Solid
   curves correspond to the MC simulation, whereas dotted curves
   correspond to the Gaussian approximation. For the MC we assume a
   true normal mass ordering with ${\Delta m^2_{32}}^{\rm true} =
   2.4\times 10^{-3} \, {\rm eV}^2$, while for the fit we minimize
   with respect to ${\Delta m^2_{32}}$ including its sign.
\label{fig:delta-theta}}
\end{figure}

We consider the test statistic $\Delta\chi^2(\deltaCP,\theta_{23})$
defined in eq.~\eqref{eq:chisq2} and simulate the distribution of this
statistic for a grid of true values in
the $(\deltaCP,\theta_{23})$ plane. Then in each point in this plane the
observed value of $\Delta\chi^2(\deltaCP,\theta_{23})$ can be compared
to the distribution from the MC to decide whether this point is
included in the confidence region at a given CL. The results of such
an analysis are shown in fig.~\ref{fig:delta-theta} for T2K only (left
panel) and the T2K + MINOS combination (right panel). The thick solid
curves indicate the confidence regions in the space of $\deltaCP$ and
$\sin^2\theta_{23}$ at $1\sigma$, $2\sigma$, and $3\sigma$ CL based on
the MC simulation. They can be compared to the thin dotted curves,
which indicate the regions obtained under the Gaussian approximation,
i.e.\ using the cut levels $t_{\chi^2}$ obtained from the $\chi^2$
distribution for 2~dof. We observe that regions obtained from the MC
are relatively close to the Gaussian limit. We conclude that the
2-dimensional test statistic has ``better'' statistical properties
than the 1-dimensional one, where the $\theta_{23}$ dependence is
profiled out.

\begin{figure}[t]
 \centering
 \includegraphics[width=\textwidth]{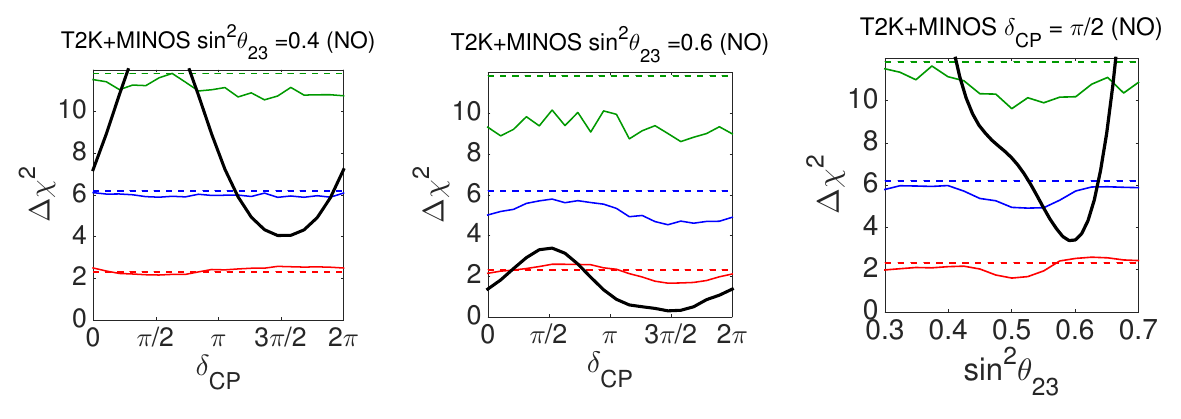}
 \mycaption{Sections through the two-dimensional
   $\Delta\chi^2(\deltaCP,\theta_{23})$ distribution for combined T2K
   and MINOS data at constant $\theta_{23}$ with $\sin^2\theta_{23} =
   0.4$ (left panel) and $\sin^2\theta_{23} = 0.6$ (middle panel), and
   at constant $\deltaCP = \pi/2$ (right panel).  Solid curves
   correspond to $t_{\rm MC}$ for $1\sigma$ (red), $2\sigma$ (blue),
   $3\sigma$ (green). Dashed lines indicate $t_{\chi^2}$ for a
   $\chi^2$ distribution with 2~dof. The black solid curves show
   $\Delta\chi^2(\deltaCP,\theta_{23})$ using the observed data.
\label{fig:2dt-sections}}
\end{figure}

This result is further illustrated in fig.~\ref{fig:2dt-sections},
which shows sections through the 2-dimensional distribution.  Those
results should not be confused with the ones shown in
figs.~\ref{fig:1d-T2K} and \ref{fig:1d-T2K+MINOS}, where the $\chi^2$
is minimised with respect to $\theta_{23}$, whereas here we keep it
fixed at the assumed true value (eq.~\eqref{eq:chisq1} versus
\eqref{eq:chisq2}). The MC curves in fig.~\ref{fig:2dt-sections}
should be compared to the corresponding cut values $t_{\chi^2}$ for a
$\chi^2$ distribution with 2~dof, indicated by the dashed lines in the
figure. We observe that the MC cut levels are close to the Gaussian
limit. For $\sin^2\theta_{23} \simeq 0.5$ we observe somewhat smaller
$t_{\rm MC}$ values compared to the Gaussian ones, due to the physical
boundary $\sin^22\theta_{23} \le 1$ (visible in the right panel in
fig.~\ref{fig:2dt-sections}). In all cases the variation of the MC cut
levels with $\deltaCP$ as well as with $\theta_{23}$ is significantly
reduced compared to the 1-dimensional case. We conclude that for
present data, the Gaussian approximation to derive confidence regions
is more reliable in the $(\deltaCP,\theta_{23})$ plane, whereas
1-dimensional confidence intervals for the CP phase $\deltaCP$ suffer
from large deviations from Gaussianity.

For figs.~\ref{fig:delta-theta} and \ref{fig:2dt-sections} we have
assumed a true normal mass ordering to generate the MC data. The
corresponding plots for a true inverted ordering are very similar.  We
can use the 2-dimensional confidence regions also to quantify the
rejection of $\deltaCP = \pi/2$ by looking for the largest CL for
which the ($\deltaCP,\sin^2\theta_{23}$) confidence regions do not
contain $\deltaCP = \pi/2$. In this way we find from the MC
calculation that combined T2K and MINOS data allow to reject $\deltaCP
= \pi/2$ at the 81.8\% (83.9\%)~CL assuming a true normal (inverted)
mass ordering. We have $\Delta\chi^2(\deltaCP=\pi/2) = 3.37$, which in
the Gaussian approximation for 2~dof corresponds to the 81.5\%~CL. We
note that the MC results for NO and IO are very similar and close to
the one obtained in the Gaussian limit. In contrast, the corresponding
rejection confidence levels based on 1-dimensional confidence
intervals for $\deltaCP$ vary strongly with the true mass ordering and
true $\theta_{23}$ and differ significantly from the Gaussian limit
(see discussion in section~\ref{sec:1d}). Note that the interpretation
of the rejection confidence levels based on 1-dimensional or
2-dimensional confidence regions is different.

\begin{figure}[t!]
  \centering
  \includegraphics[width=0.6\textwidth]{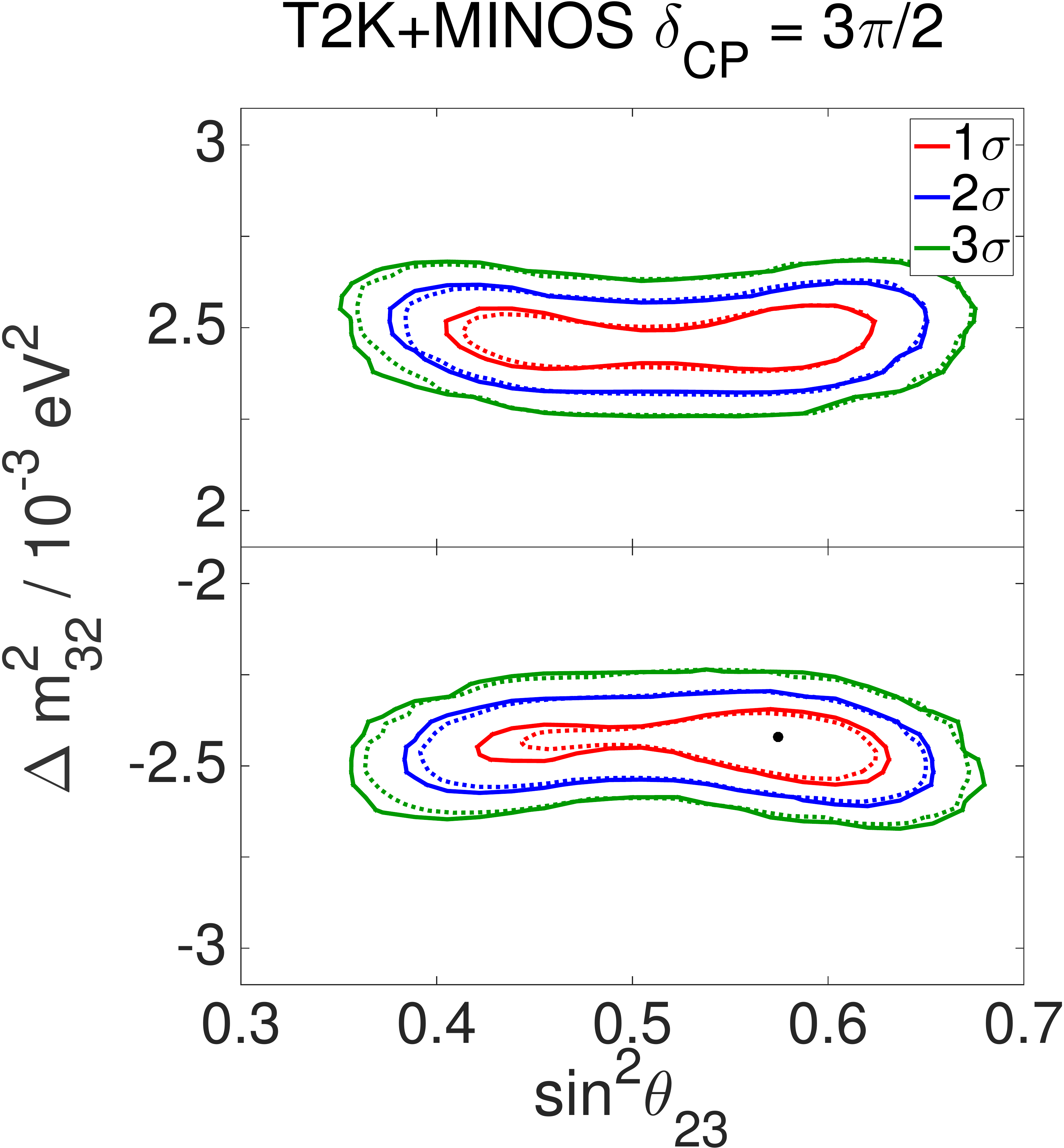}
  \mycaption{Two-dimensional confidence regions at $1\sigma$ (red),
    $2\sigma$ (blue), $3\sigma$ (green) in the $(\theta_{23}, \Delta
    m^2_{32})$ plane for T2K + MINOS data. Solid curves correspond to
    the MC simulation, whereas dotted curves correspond to the
    Gaussian approximation. For the MC we assume a true value of
    $\deltaCP = 3\pi/2$, while for the fit we minimize with respect to
    $\deltaCP$.
\label{fig:theta-Dmq}}
\end{figure}

In Fig.~\ref{fig:theta-Dmq} we show the 2-dimensional Feldman-Cousins
confidence regions in the plane of $\sin^2\theta_{23}$ and $\Delta
m^2_{32}$ for the combined T2K and MINOS data. We observe that they
agree quite well with the standard Gaussian approximation. For
generating the MC data we have assumed here $\deltaCP = 3\pi/2$, but
the results are very similar for other true values of $\deltaCP$.

\section{Increased exposure and T2K anti-neutrino data}
\label{sec:future}

Next we are going to discuss how this situation will change in the
near future, for increased exposure in T2K or when data on
anti-neutrinos become available. Following the T2K
collaboration~\cite{Abe:2014tzr} we consider an exposure of $7.8\times
10^{21}$ protons-on-target (p.o.t.), approximately a factor 12 larger
than the current exposure.  We consider two cases, either using all of
this exposure for neutrino data or equally sharing the exposure
between neutrino and anti-neutrino running. We depart from the code
used for our present-data T2K analysis (see section~\ref{sec:data}),
and scale the spectrum normalization such that we can reproduce the
expected number of events given in tables~4 and 5
in~\cite{Abe:2014tzr}. This is a rough approximation, especially for
the anti-neutrino case, where we ignore the (rather substantial)
contribution from the neutrino component in the beam. Despite those
simplifications we can reproduce accurately the event spectra from
fig.~2 in~\cite{Abe:2014tzr}, as well as the sensitivity plots based
on Asimov data given in ref.~\cite{Abe:2014tzr}. Note that total event
numbers are still relatively small, 210 (260) events for neutrinos and
49 (35) events for antineutrinos for $\deltaCP = 0$ ($3\pi/2$),
implying large statistical errors. Our approximate implementation
suffices to study the expected statistical properties of the test
statistics. A detailed and accurate sensitivity calculation is beyond
the scope of this work.

\begin{figure}[t!]
\includegraphics[width=\textwidth]{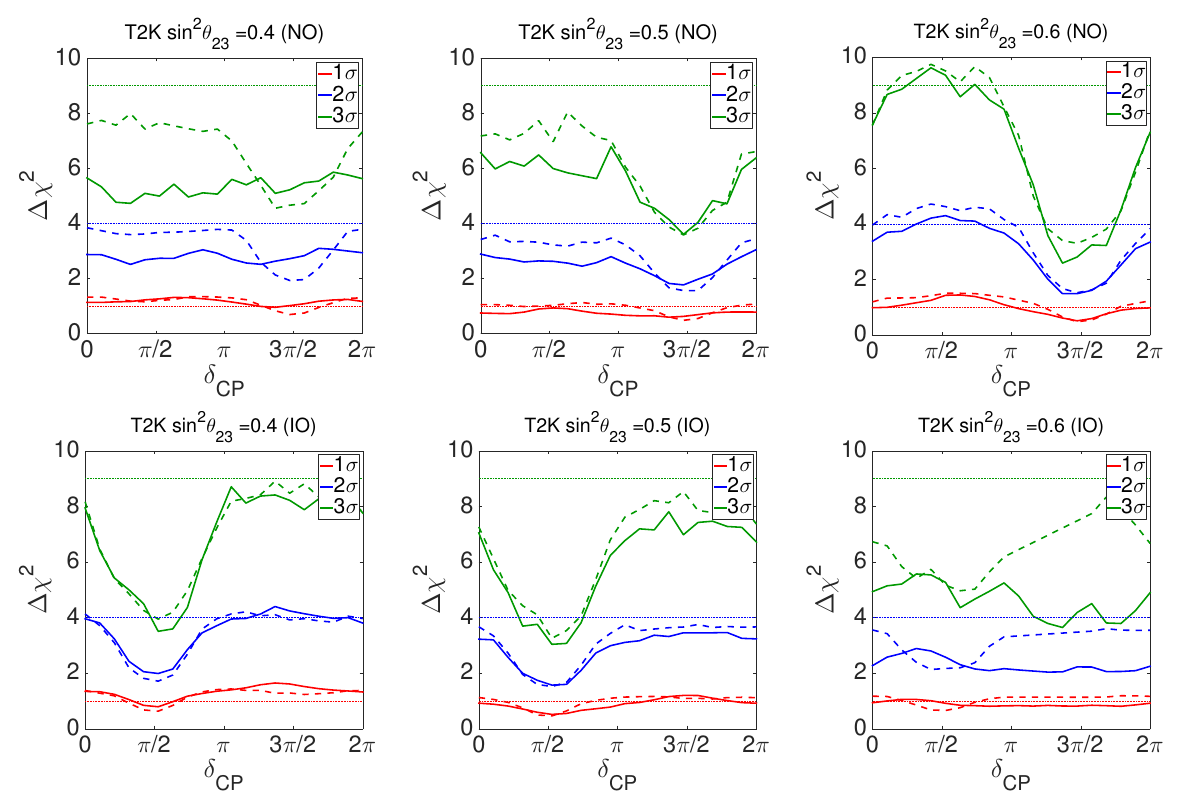}
\mycaption{The cut levels $t_{\rm MC}({\rm CL}, \Theta)$ for
  $\Delta\chi^2(\deltaCP)$ for future T2K data of $7.8\times
  10^{21}$~p.o.t.\ for $1\sigma$ (red), $2\sigma$ (blue), $3\sigma$
  (green). Solid curves assume 100\% neutrino running, dashed curves
  are for 50\% neutrino and anti-neutrino data, each.  Left, middle,
  right panels correspond to $\sin^2\theta_{23}^{\rm true} = 0.4, 0.5,
  0.6$, respectively. We take $|{\Delta m^2_{32}}^{\rm true}| =
  2.4\times 10^{-3} \, {\rm eV}^2$ and for the upper (lower) row we
  have assumed a true normal (inverted) mass ordering. The mass
  ordering ambiguity is included in the minimisation.}
\label{fig:future-all}
\end{figure}

In fig.~\ref{fig:future-all} we show the cut levels for
$\Delta\chi^2(\deltaCP)$ for different assumptions about the true
$\theta_{23}$ and mass ordering, both for neutrino-only data and for
combining neutrino and anti-neutrino data. We find that in both cases
a significant dependence on the unknown true values of $\theta_{23}$
and the mass ordering remains. In particular, the significance of
excluding values of $\deltaCP \simeq \pi/2$ or $3\pi/2$ will vary
quite strongly. The locations of the dips in the cut levels follow the
pattern discussed in section~\ref{sec:probab}.  The regions of low cut
levels for neutrino data-only visible for NO, $\sin^2\theta_{23} =
0.6$, $\deltaCP \simeq 3\pi/2$ (upper right panel) and IO,
$\sin^2\theta_{23} = 0.4$, $\deltaCP \simeq \pi/2$ (lower left panel)
correspond to the regions of maximal and minimal oscillation
probability indicated in the first two lines of eq.~\eqref{eq:Pminmax}. Also the dips in the
middle panels of fig.~\ref{fig:future-all} (for $\sin^2\theta_{23} =
0.5$) follow this argument.

\begin{figure}[t!]
\includegraphics[width=\textwidth]{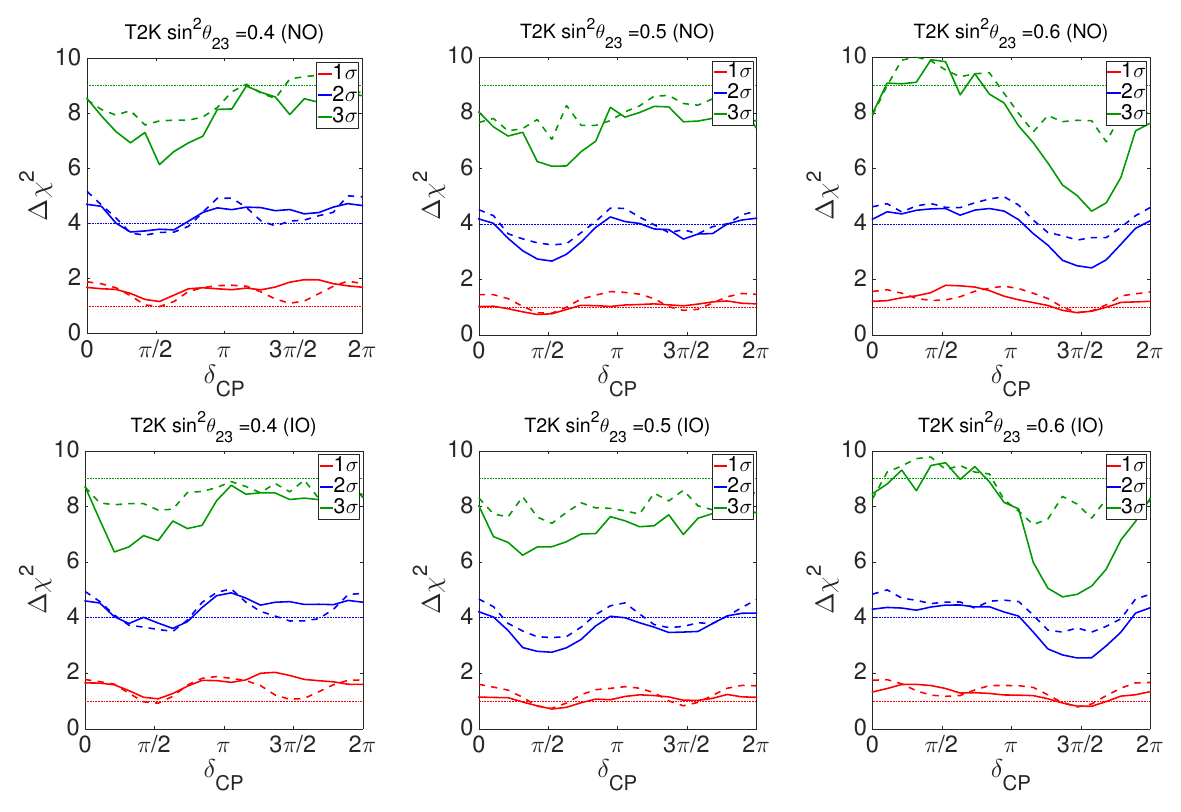}
\mycaption{Same as figure~\ref{fig:future-all}, but assuming that the
  mass ordering is known, i.e.\ restricting the fit to the assumed
  true ordering.}
\label{fig:future-trueMO}
\end{figure}

We also observe from the figure that in many cases using anti-neutrino
data has only a small impact on the distribution of the test
statistics. The only exceptions are $\sin^2\theta_{23} = 0.4$, NO
(upper left) and $\sin^2\theta_{23} = 0.6$, IO (lower right).  In
those cases cut levels are raised close to the Gaussian limit for $0
\le \deltaCP \le \pi$ or $\pi \le \deltaCP \le 2\pi$, respectively.
In the opposite regions of $\deltaCP$ for those cases cut levels remain
low also when anti-neutrino data are added.  For those regions, there
is an octant degenerate solution which basically destroys any
sensitivity to $\deltaCP$, see panel (d) in
fig.~\ref{fig:sensitivity}, while adding information from
anti-neutrinos significantly helps in resolving this degeneracy which
increases the sensitivity to $\deltaCP$.  Those regions correspond to
the minimal and maximal values for the anti-neutrino oscillation
probability indicated in the last two lines of
eq.~\eqref{eq:Pminmax}, explaining the good sensitivity of
anti-neutrino data for those cases.

In fig.~\ref{fig:future-trueMO} we show the same analysis as in
fig.~\ref{fig:future-all}, except that we assume that the mass
ordering is known (determined independently by some other experiment).
Hence, we do not minimize with respect to the mass ordering, but
restrict the fit to the assumed true ordering. We observe that the
sign($\Delta m^2_{32}$) degeneracy has a large impact on the
distribution of the test statistics and cut levels become much closer
to the Gaussian approximation.\footnote{This is an example where the
  presence of the degeneracy decreases the effective number of dof,
  contrary to the cases discussed previously, where the presence of a
  degeneracy increases the number of dof.} This is different to the
situation we find for current data discussed in
section~\ref{sec:results}, where the mass ordering degeneracy has only
a small impact on the distribution of the test statistics. For
neutrino-only data, there are still a few cases of large deviations
from Gaussianity, whereas the combined neutrino and anti-neutrino data
lead to cut levels close to the $\chi^2$ limit (see dashed curves in
fig.~\ref{fig:future-trueMO}). We still observe small dips for
$\deltaCP \simeq \pi/2$ and $3\pi/2$ (now pretty symmetric around
$\deltaCP=\pi$), which can be traced back to the $\sin\deltaCP$
dependence of the probabilities, leading to a reduction of the dof for
those values of $\deltaCP$. Note also that in some cases we find now
cut levels which are even higher than the Gaussian ones. A possible
explanation for this behavior can be found in the discussion in
section~\ref{sec:probab}, see item~1.

\section{Discussion and conclusions}
\label{sec:conclusions} 

We have studied in detail the information we can obtain on the
leptonic CP phase $\deltaCP$ from current data, focusing on the
robustness of frequentist confidence regions, by performing a Monte
Carlo simulation of the data from the T2K and MINOS experiments. We
attempt to quantify the current preference for $\deltaCP \simeq
3\pi/2$ over $\deltaCP \simeq \pi/2$. We have focused on the interplay
of the main unknown parameters, namely $\deltaCP$, $\theta_{23}$, and
the neutrino mass ordering.  Our findings can be summarized as
follows.

The distribution of the $\Delta\chi^2$ test statistic used for
1-dimensional confidence intervals for $\deltaCP$ shows large
deviations from the Gaussian limit. In particular, it strongly depends
on the unknown true value of $\theta_{23}$. This introduces an
ambiguity in the confidence intervals for $\deltaCP$, see
fig.~\ref{fig:deltaCIs}. While the $1\sigma$ interval for $\deltaCP$
is relatively stable and close to the Gaussian approximation, at
higher confidence level large variations occur. In particular, the CL
with which values of $\deltaCP \simeq \pi/2$ are disfavored ranges
from $2\sigma$ to $3\sigma$, depending on $\theta_{23}$ and the mass
ordering. We can trace back the origin of those results to the
complicated non-linear parameter dependence of the relevant
oscillation probabilities (trigonometric dependence of $\deltaCP$ and
$\theta_{23}$-octant and mass ordering degeneracies), combined with
the rather poor sensitivity of current data to $\deltaCP$.

We conclude that one should not use the Gaussian approximation when
making statements about $\deltaCP$ based on the 1-dimensional
$\Delta\chi^2$ test statistic. Typically the ``true'' confidence
levels obtained from the Monte Carlo simulation lead to more
restrictive confidence intervals and to stronger rejections of values
of $\deltaCP$ around $\pi/2$. In this sense the use of the Gaussian
approximation is conservative. We have shown that the Gaussian
approximation is better justified for 2-dimensional confidence regions
in the plane of $\deltaCP$ and $\sin^2\theta_{23}$, see
fig.~\ref{fig:delta-theta}. In particular, the dependence of the
$\Delta\chi^2$ distribution on the true values of $\deltaCP$ and
$\theta_{23}$ is much less severe in the 2-dimensional case.  The
2-dimensional confidence region in the $(\deltaCP, \theta_{23})$ plane
for combined T2K and MINOS data does not include $\deltaCP = \pi/2$ up
to the 81.8\% (83.9\%)~CL assuming a true normal (inverted) mass
ordering.\footnote{When first data \cite{T2K-EPS15} from T2K on
  $\overline\nu_\mu\to \overline\nu_e$ are included, the corresponding
  numbers are 86.3\% (89.2\%)~CL, see appendix.} Those values are
close to the CL of 81.5\% obtained under the Gaussian approximation.

We have considered also the 1-dimensional confidence intervals for
$\theta_{23}$ and $\Delta m^2_{32}$. For $\theta_{23}$ we find
approximately Gaussian behavior, with some deviations around maximal
mixing, see fig.~\ref{fig:1theta}. This is a manifestation of the
boundary $\sin^22\theta_{23} \le 1$, which implies that the derivative
of the event rates predicted for the disappearance channel with
respect to $\theta_{23}$ is zero for $\theta_{23} = 45^\circ$. The
test statistic for $\Delta m^2_{32}$ has a distribution very close to
the Gaussian limit, as well as the 2-dimensional confidence regions in
the $(\sin^2\theta_{23}, \Delta m^2_{32})$ plane.

In section~\ref{sec:future} we have studied the distribution of the
1-dimensional $\Delta\chi^2$ test statistic for $\deltaCP$ assuming an
increased exposure for T2K of $7.8\times 10^{21}$ protons-on-target,
roughly a factor 12 larger than current exposure, where we consider
also the possibility of using half of this exposure for anti-neutrino
running. We find that even in this case large deviations from the
Gaussian behavior can be expected. Typically reduced cut
levels for the $\Delta\chi^2$ are obtained around either $\deltaCP \simeq \pi/2$ or
$3\pi/2$, depending on the unknown true value of
$\theta_{23}$ and the mass ordering. Close to Gaussian behavior is
only obtained for neutrino plus anti-neutrino running and assuming
that the neutrino mass ordering is known. 

Let us mention that in the global fit of all oscillation data also
SuperKamiokande atmospheric neutrino data contribute to the
determination of $\theta_{23}$ and to a small extent also of
$\deltaCP$, see ref.~\cite{Gonzalez-Garcia:2014bfa} for a
discussion. Ideally a combined MC simulation of long-baseline and
atmospheric neutrino data should be performed, which however, is not
feasible due to the numerical complexity of the atmospheric neutrino
fit. Since atmospheric neutrino data play only a subleading role for
$\deltaCP$ we expect that the results presented here would not be
modified substantially by including atmospheric neutrinos.

For the investigation of near term future data in
section~\ref{sec:future} we have not considered the NOvA
experiment~\cite{Ayres:2004js, Patterson:2012zs}, from which data will
become available during the next years. In general we expect improved
behaviour of the relevant test statistics, since complementary data
from NOvA may help to resolve some of the degeneracies (see,
e.g.\ refs.~\cite{Machado:2013kya, Ghosh:2015ena} for recent studies),
which---as we have shown---play a crucial role for the deviations from
Gaussianity. The same is true also for experiments aiming to determine
the neutrino mass ordering, see ref.~\cite{Blennow:2013oma} for a
discussion and references. An exhaustive investigation of the expected
statistical properties of future data is beyond the scope of this
work. Some discussion along those lines in the context of CP violation
can be found in ref.~\cite{Blennow:2014sja}. It will be an interesting
topic for future work to study sensitivities to $\deltaCP$ of combined
data from NOvA, T2K, and other upcoming experiments based on the true
distributions of the relevant test statistics.

\subsection*{Acknowledgements}

We are grateful to Michele Maltoni for sharing with us the NuFit
long-baseline code. We thank Mattias Blennow for useful discussions
and Pilar Coloma for comments on the manuscript. We acknowledge
support from the European Union FP7 ITN INVISIBLES (Marie Curie
Actions, PITN-GA-2011-289442).

\appendix

\section{Impact of first anti-neutrino data from T2K}
\label{sec:app}

After completion and submission of this paper, the first anti-neutrino
results from T2K were presented at the EPS HEP 2015 conference
\cite{T2K-EPS15}. In this appendix we show the impact of those data on
the determination of $\deltaCP$.

The results presented in \cite{T2K-EPS15} corresponds to about
$4\times 10^{20}$~p.o.t.\ in the anti-neutrino mode with 3 observed
events in the appearance channel.\footnote{We do not consider the
  disappearance channel here.}  The expected background (NC and other)
is 1.17 events, while the predicted signal from $\overline\nu_\mu \to
\overline\nu_e$ ($\nu_\mu \to \nu_e$) induced events ranges from 2 to 4
(0.3 to 0.6) events, depending on $\deltaCP$ and the mass ordering. We
could reproduce the predicted number of events given in
\cite{T2K-EPS15} with good accuracy, which allows us to include those
data and combine it with the other data used in this work. We use a
$\chi^2$ as given in eq.~\eqref{eq:chisq} with just one bin (only the
total number of events is fitted), taking into account the background
expectation, as well as oscillated anti-neutrino and neutrino event
predictions.

Clearly the statistical significance of those results is poor, since 3
observed events are even consistent with the background only
hypothesis of about 1.17 events (no oscillation induced events at all)
at slightly more than $1\sigma$. Therefore we expect that those data
will change $\Delta\chi^2$ by about 1 unit. Nevertheless,
anti-neutrinos carry complementary information to the neutrino data
and therefore it may be interesting to investigate the impact of those
results.

\begin{figure}[t!]
 \centering
 \includegraphics[width=\textwidth]{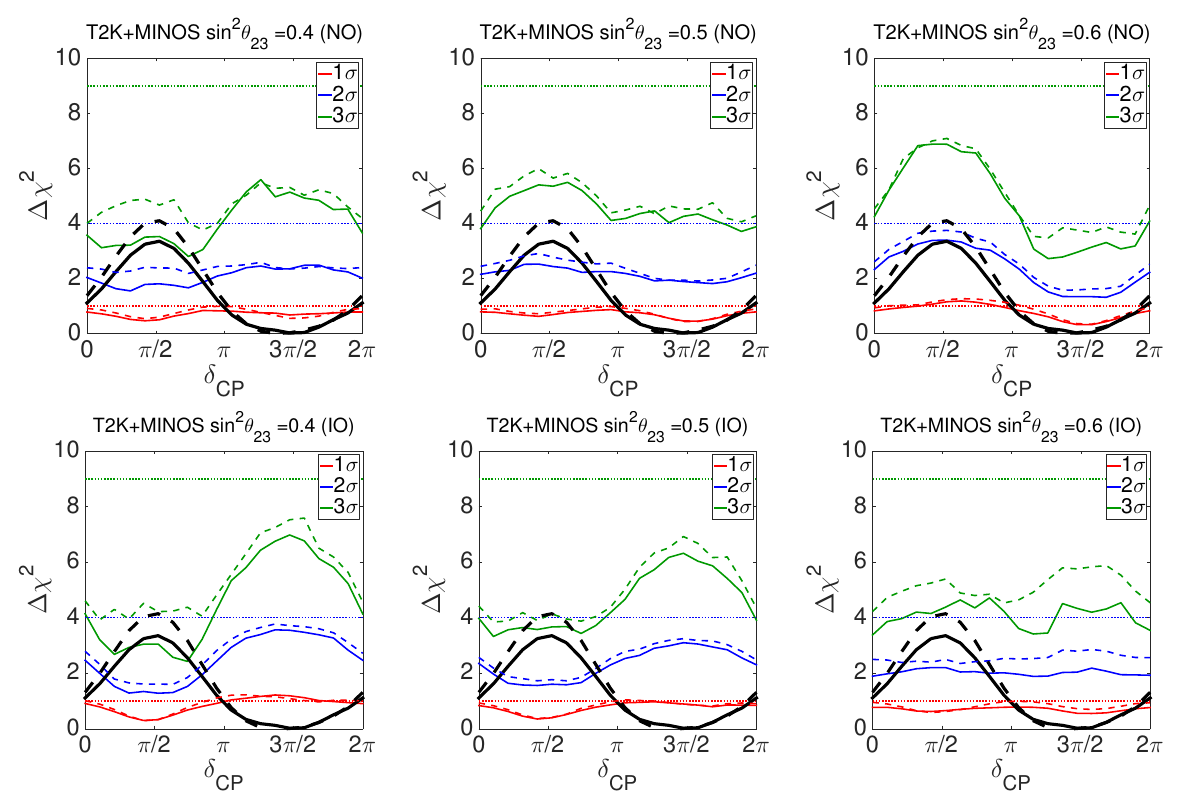}
 \mycaption{The cut levels $t_{\rm MC}({\rm CL}, \Theta)$ for
   $\Delta\chi^2(\deltaCP)$ from combined T2K and MINOS data for
   $1\sigma$ (red), $2\sigma$ (blue), $3\sigma$ (green). Left, middle,
   right panels correspond to $\sin^2\theta_{23}^{\rm true} = 0.4,
   0.5, 0.6$, respectively. We take $|{\Delta m^2_{32}}^{\rm true}| =
   2.4\times 10^{-3} \, {\rm eV}^2$ and for the upper (lower) row we
   have assumed a true normal (inverted) mass ordering. The black
   curves show $\Delta\chi^2(\deltaCP)$ using the observed data (same
   curves in all panels). Solid curves correspond to the data given in
   tab.~\ref{tab:data} (no T2K anti-neutrinos, same as in
   fig.~\ref{fig:1d-T2K+MINOS}), dashed curves include T2K
   anti-neutrino data. \label{fig:1d-T2K+MINOS_anti}}
\end{figure}

\begin{figure}[t!]
  \includegraphics[width=\textwidth]{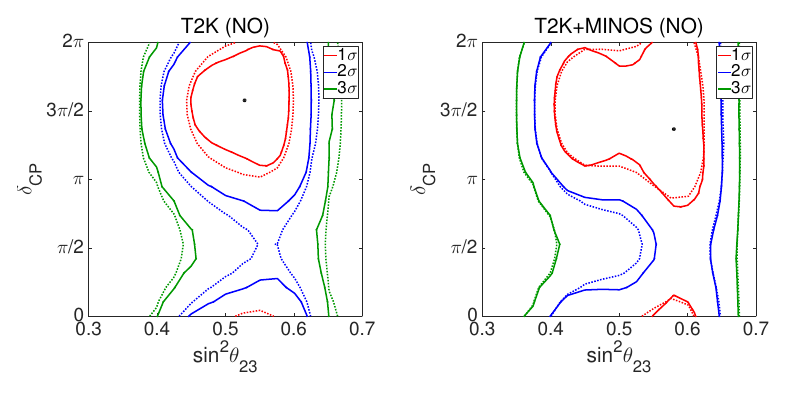}
  \mycaption{Two-dimensional confidence regions at $1\sigma$ (red),
    $2\sigma$ (blue), $3\sigma$ (green) in the $(\sin^2\theta_{23},
    \deltaCP)$ plane for T2K (left panel) and T2K + MINOS (right
    panel), including T2K anti-neutrino data in both cases. Solid
    curves correspond to the MC simulation, whereas dotted curves
    correspond to the Gaussian approximation. For the MC we assume a
    true normal mass ordering with ${\Delta m^2_{32}}^{\rm true} =
    2.4\times 10^{-3} \, {\rm eV}^2$, while for the fit we minimize
    with respect to ${\Delta m^2_{32}}$ including its sign.
\label{fig:delta-theta_anti}}
\end{figure}

Fig.~\ref{fig:1d-T2K+MINOS_anti} shows the impact of the anti-neutrino
data on the 1-dimensional confidence interval for $\deltaCP$. The
solid curves, both for the cut levels as well as for
$\Delta\chi^2(\deltaCP)$ are without T2K anti-neutrinos and are
identical to fig.~\ref{fig:1d-T2K+MINOS}, while the dashed curves
include the information from the T2K anti-neutrino events. Comparing
the black solid and dashed curve we find that $\Delta\chi^2$ for
$\deltaCP = \pi/2$ is increased by about 0.75. However,
also the cut levels are increased by a similar amount and hence the
significance of rejecting $\deltaCP=\pi/2$ is hardly affected. For
$\sin^2\theta_{23} = 0.4$ the effect is most pronounced and in those
cases the significance actually decreases, despite the increased
$\Delta\chi^2$.

In fig.~\ref{fig:delta-theta_anti} we show the 2-dimensional
confidence region in the $(\deltaCP,\theta_{23})$ plane for T2K as
well as T2K+MINOS data, including the T2K anti-neutrino events in both
cases. This figure should be compared with fig.~\ref{fig:delta-theta},
which shows the corresponding regions without T2K anti-neutrinos. As
expected the difference is small, with the size of the confidence
region being slightly decreased due to the new data. Including T2K
anti-neutrinos we find that combined T2K and MINOS data allow to
reject $\deltaCP = \pi/2$ at the 86.3\% (89.2\%)~CL assuming a true
normal (inverted) mass ordering, to be compared with 81.8\%
(83.9\%)~CL without T2K anti-neutrinos.  For
$\Delta\chi^2(\deltaCP=\pi/2)$ we find now a value of 4.27, which in
the Gaussian approximation for 2~dof corresponds to the 88.2\%~CL.

\bibliographystyle{my-h-physrev}
\bibliography{./refs}

\end{document}